\documentclass{aa}
\usepackage{graphicx}
\usepackage{txfonts}
\usepackage{epsfig}
\usepackage{natbib}

\def\PLUTO{{\sc pluto}}

\newcommand\rs[1]{_\mathrm{#1}}

\begin{document} 
\title{3D MHD modeling of the expanding remnant of SN\,1987A}
\subtitle{Role of magnetic field and non-thermal radio emission}

\author{S.\ Orlando\inst{1}
   \and M.\ Miceli\inst{2,1}
   \and O.\ Petruk\inst{3,4}
   \and M.\ Ono\inst{5}
   \and S.\ Nagataki\inst{5,6}
   \and M. A.\ Aloy\inst{7}
   \and P.\ Mimica\inst{7}
   \and \\ S.-H.\ Lee\inst{8}
   \and F.\ Bocchino\inst{1}
   \and G.\ Peres\inst{2,1}
   \and M.\ Guarrasi\inst{9}
}

\offprints{S. Orlando}

\institute{INAF -- Osservatorio Astronomico di Palermo, Piazza del Parlamento 1, I-90134 Palermo, Italy\\ 
\email{salvatore.orlando@inaf.it}
\and Dip. di Fisica e Chimica, Universit\`a degli Studi di Palermo, Piazza del Parlamento 1, 90134 Palermo, Italy
\and Institute for Applied Problems in Mechanics and Mathematics, Naukova Street, 3-b Lviv 79060, Ukraine
\and Astronomical Observatory of the Jagiellonian University, ul. Orla 171, 30-244 Krak\'ow, Poland
\and Astrophysical Big Bang Laboratory, RIKEN Cluster for Pioneering Research, 2-1 Hirosawa, Wako, Saitama 351-0198, Japan
\and RIKEN Interdisciplinary Theoretical \& Mathematical Science Program (iTHEMS), 2-1 Hirosawa, Wako, Saitama 351-0198, Japan 
\and Dep. de Astronom\'ia y Astrof\'isica, Univ. de Valencia, Ed. de Investigaci\'on Jeroni Munyoz, E-46100 Burjassot, Valencia, Spain
\and Kyoto University, Department of Astronomy, Oiwake-cho, Kitashirakawa, Sakyo-ku, Kyoto 606-8502, Japan
\and CINECA-Interuniversity consortium, via Magnanelli 6/3, I-40033, Casalecchio di Reno, Bologna, Italy
  }

\date{Received date / Accepted date}

\abstract
{}
{We investigate the role played by a pre-supernova (SN) ambient magnetic
field on the dynamics of the expanding remnant of SN\,1987A and the origin
and evolution of the radio emission from the remnant, in particular,
during the interaction of the blast wave with the nebula surrounding
the SN.}
{We model the evolution of SN\,1987A from the breakout of the shock
wave at the stellar surface to the expansion of its remnant through
the surrounding nebula by three-dimensional magnetohydrodynamic
simulations. The model considers the radiative cooling, the deviations
from equilibrium of ionization, the deviation from
temperature-equilibration between electrons and ions, and a plausible
configuration of the pre-SN ambient magnetic field. We explore
strengths of the pre-SN magnetic field ranging between $1 \mu$G and
$100 \mu$G at the inner edge of the nebula and we assume an
average field strength at the stellar surface $B_0 \approx 3$~kG.
From the simulations, we synthesize both thermal X-ray and non-thermal
radio emission and compare the model results with observations.}
{The presence of an ambient magnetic field with strength in the range
considered does not change significantly the overall evolution of
the remnant. Nevertheless, the magnetic field reduces the erosion and
fragmentation of the dense equatorial ring after the impact of the SN
blast wave. As a result, the ring survives the passage of the blast, at
least, during the time covered by the simulations (40 years). Our model
is able to reproduce the morphology and lightcurves of SN\,1987A in both
X-ray and radio bands. The model reproduces the observed radio emission
if the flux originating from the reverse shock is heavily suppressed. In
this case, the radio emission originates mostly from the forward shock
traveling through the H\,II region and this may explain why the radio
emission seems to be insensitive to the interaction of the blast with the
ring. Possible mechanisms for the suppression of emission from the
reverse shock are investigated. We find that synchrotron self-absorption
and free-free absorption have negligible effects on the emission during
the interaction with the nebula. We suggest that the emission from the
reverse shock at radio frequencies might be limited
by highly magnetized ejecta.}
{}


\titlerunning{3D MHD modeling of the expanding remnant of SN\,1987A}
\authorrunning{S. Orlando et~al.}

\maketitle

\section{Introduction}
\label{sec:intro}

The expanding remnant of SN\,1987A in the Large Magellanic Cloud
offers the opportunity to study in great detail the transition from
the phase of supernova (SN) to that of supernova remnant (SNR),
thanks to its youth (the SN was observed on 1987 February 23;
\citealt{1987A&A...177L...1W}) and proximity (at about 51.4~kpc from
Earth; \citealt{1999IAUS..190..549P}). SN\,1987A has been the subject of
an intensive international observing campaign, incorporating observations
at every possible wavelength. The SN explosion and the subsequent
evolution of its remnant have been monitored continuously since the
collapse of the progenitor star, Sanduleak (Sk) $−69^{\rm o} 202$,
a blue supergiant (BSG) with an initial mass of $\approx 20\,M_{\odot}$
(\citealt{1987Natur.327..597H}).

Observations in different wavelength bands have revealed that the
pre-SN circumstellar medium (CSM) around SN\,1987A is highly
inhomogeneous (e.g. \citealt{1989ApJ...347L..61C, 2005ApJS..159...60S}).
It is characterized by an extended nebula consisting mainly of three dense
rings, one lying in the equatorial plane, and two less dense ones lying
in planes almost parallel to the equatorial one, but displaced by
about 0.4 pc above and below the central ring; the rings are immersed
in a much larger and tenuous H\,II region with peanut-shaped structure
with a maximum extension of about 6~pc in the direction perpendicular
to the equatorial plane. General consensus is that the stellar
progenitor went, first, through a phase of red supergiant (RSG)
and, then, through that of BSG before the collapse
(\citealt{2000ApJ...528..426C}). The nebula resulted from the
interaction between the fast wind emitted during the phase of BSG and
the relic denser and slower wind emitted previously during the phase
of RSG (e.g. \citealt{1995ApJ...452L..45C, 2007Sci...315.1103M, 2008A&A...488L..37C}).

Radio and X-ray observations collected in the last $\approx 30$ years
have monitored in detail the interaction of the SN blast wave with the
nebula (for a comprehensive review see \citealt{2016ARA&A..54...19M}). The
interaction started about 3~years after the explosion, when both radio and
X-ray fluxes have suddenly increased (e.g. \citealt{1996A&A...312L...9H,
1997ApJ...479..845G}). This was interpreted as being due to the
blast wave impacting onto the H\,II region and starting to travel in
a medium much denser than that of the BSG wind. In the subsequent
years the radio and X-ray fluxes have continued to rise together.
After about 14~years, however, the soft X-ray light curve has steepened
still further, contrary to the hard X-ray and radio lightcurves
(\citealt{2005ApJ...634L..73P}) which continued to increase with
almost constant slope. This has indicated a source of emission that
leads predominantly to soft X-rays and was interpreted as being
due to the blast wave sweeping up the dense central equatorial ring
(\citealt{2007AIPC..937....3M}). Since then the soft X-ray lightcurve
has continued to increase indicating that the blast wave was traveling
through the dense ring (\citealt{2013ApJ...764...11H}). In its latest
X-ray observations, {\it Chandra} has recorded a significant change
in the soft X-ray lightcurve which has remained constant since year 26
(\citealt{2016ApJ...829...40F}). This has been interpreted as being
due to the blast wave leaving the ring and starting to travel in a
less dense environment. A similar conclusion has been obtained
more recently in the radio band from the analysis of observations
taken with the Australia Telescope Compact Array (ATCA;
\citealt{2018arXiv180902364C}).

The evolution and morphology of the radio emission from SN\,1987A
have been investigated by \cite{2014ApJ...794..174P} through a
three-dimensional (3D) hydrodynamic (HD) model describing the
interaction of the SN blast wave with the nebula between years 2
and 27. Their best-fit model is able to reproduce the evolution of
the radio emission reasonably well, although it predicts a significant
steepening of radio flux when the blast wave hits the central ring,
at odds with observations. As for the evolution of the X-ray emission,
\cite{2015ApJ...810..168O} have proposed a 3D HD model which describes
the evolution of SN\,1987A since the shock breakout at the stellar
surface (few hours after the core-collapse), and which covers 40
years of evolution. Their best-fit model is able to reproduce quite well,
at the same time, the bolometric lightcurve of the SN observed during
the first 250 days of evolution, and the X-ray emission (morphology,
lightcurves, and spectra) of the subsequent expanding remnant (see also
Miceli et al. 2018, submitted to Nature Astronomy).

Current models of SN\,1987A, however, do not include the effects
of an ambient magnetic field. The latter is expected to not affect
significantly the overall evolution (and the morphology) of the
remnant. Nevertheless it may play a role during the interaction of
the blast wave with the ring by dumping the HD instabilities
(responsible for the fragmentation of the ring) that would develop
during the interaction (e.g. \citealt{2008ApJ...678..274O}).
Also, modeling the magnetic field is necessary
to synthesize self-consistently from the models the non-thermal
radio emission. In the model of \cite{2014ApJ...794..174P}
the magnetic field is not modeled but the radio emission is synthesized
by assuming a randomly oriented magnetic field at the shock front,
using analytic estimates for cosmic-ray (CR) driven magnetic field
amplification. Also, it is not clear from the literature where
the radio emission originates and why the radio flux seems to be
insensitive to the interaction of the blast wave with the dense
ring.

Here we investigate the role played by an ambient magnetic field
on the evolution of the remnant and, more specifically, during the
interaction of the SN blast wave with the nebula. Also, we aim at
exploring the origin of non-thermal radio emission from SN\,1987A
and possible diagnostics of the inhomogeneous CSM in which the
explosion occurred. To this end, we model the evolution of SN\,1987A
using detailed 3D MHD simulations. From the simulations we synthesize
the thermal X-ray and non-thermal radio emission (including the
modeling of the post-shock evolution of relativistic electrons and
the observed evolution of the radio spectral index) and compare the
synthetic lightcurves with those inferred from the analysis of
observations.

In Sect. \ref{sec:model} we describe the MHD model, the numerical
setup, and the synthesis of thermal X-ray and non-thermal radio
emission; in Sect. \ref{sec:result} we discuss the results; and
finally in Sect. \ref{sec:conc} we draw our conclusions.

\section{MHD modeling}
\label{sec:model}

The model describes the evolution of SN\,1987A from the breakout of
the shock wave at the stellar surface (occurring few hours after the
SN event) to the interaction of the blast wave and ejecta caused
by the explosion with the surrounding nebula. The model covers 40
years of evolution to provide also some hints on the evolution of
the expanding remnant expected in the near future, namely when the
Square Kilometer Array (SKA) in radio and the Athena satellite in
X-rays will be in operation. The computational strategy is the same as
described in \cite{2015ApJ...810..168O} and consists of two steps (see
also \citealt{2016ApJ...822...22O}): first, we modeled the ``early''
post-explosion evolution of a CC-SN during the first 24 hours through
1D simulations; then, we mapped the output of these simulations into 3D
and modeled the transition from the phase of SN to that of SNR and the
interaction of the remnant with the inhomogeneous pre-SN environment
(see \citealt{2015ApJ...810..168O} for further details).

In this work, for the post-explosion evolution of the SN during the
first 24 hours, we adopted the best-fit model (model SN-M17-E1.2-N8)
described in \cite{2015ApJ...810..168O} (see their Table~3); the
model is characterized by an explosion energy of $1.2\times
10^{51}$~erg and an envelope mass of 17~$M_{\odot}$. The SN evolution 
has been simulated by a 1D Lagrangian code in spherical geometry
which solves the equations of relativistic radiation hydrodynamics,
for a self-gravitating matter fluid interacting with radiation
(\citealt{2011ApJ...741...41P}). The code is fully general relativistic
and provides an accurate treatment of radiative transfer at all
regimes (from the one in which the ejecta are optically thick up
to when they are optically thin). 

This 1D simulation of the SN provides the radial distribution of
ejecta (density, pressure, and velocity) about 24 hours after the
SN event. Then we mapped the 1D profiles in the 3D domain and started
3D MHD simulations which describe the interaction of the remnant
with the surrounding nebula. The evolution of the blast wave was
modeled by numerically solving the full time-dependent MHD equations
in a 3D Cartesian coordinate system $(x, y, z)$, including the
effects of the radiative losses from optically thin plasma.
The MHD equations were solved in the non-dimensional conservative
form:

\begin{equation}
\frac{\partial \rho}{\partial t} + \nabla \cdot (\rho \vec{u}) = 0~,
\end{equation}

\begin{equation}
\frac{\partial \rho \vec{u}}{\partial t} + \nabla \cdot (\rho
\vec{u}\vec{u}-\vec{B}\vec{B}) + \nabla P_* = 0~,
\end{equation}

\begin{equation}
\frac{\partial \rho E}{\partial t} +\nabla\cdot [\vec{u}(\rho
E+P_*) -\vec{B}(\vec{u}\cdot \vec{B})] = -n_{\rm e} n_{\rm H}
\Lambda(T)~,
\end{equation}

\begin{equation}
\frac{\partial \vec{B}}{\partial t} +\nabla
\cdot(\vec{u}\vec{B}-\vec{B}\vec{u}) = 0~,
\end{equation}

\noindent
where

\[
P_* = P + \frac{B^2}{2}~,~~~~~~~~~~~~~
E = \epsilon +\frac{1}{2} u^2+\frac{1}{2}\frac{B^2}{\rho}~,
\]

\noindent
are the total pressure, and the total gas energy (internal energy,
$\epsilon$, kinetic energy, and magnetic energy) respectively, $t$
is the time, $\rho = \mu m_H n_{\rm H}$ is the mass density, $\mu$
is the mean atomic mass, $m_H$ is the mass of the hydrogen atom,
$n_{\rm H}$ is the hydrogen number density, $\vec{u}$ is the gas
velocity, $T$ is the temperature, $\vec{B}$ is the magnetic field,
and $\Lambda(T)$ represents the optically thin radiative losses per
unit emission measure derived with the PINTofALE spectral code
\citep{Kashyap2000BASI} with the APED V1.3 atomic line database
\citep{2001ApJ...556L..91S}, assuming metal abundances appropriate for SN
1987A (\citealt{2009ApJ...692.1190Z}). We used the ideal gas law,
$P=(\gamma-1) \rho \epsilon$, where $\gamma=5/3$ is the adiabatic
index.

The simulations of the expanding SNR were performed using \PLUTO\
\citep{2007ApJS..170..228M, 2012ApJS..198....7M}, a modular
Godunov-type code intended mainly for astrophysical applications
and high Mach number flows in multiple spatial dimensions. The code
is designed to make efficient use of massively parallel computers
using the message-passing interface (MPI) for interprocessor
communications. The MHD equations are solved using the MHD module
available in \PLUTO, configured to compute intercell fluxes with
the Harten-Lax-van Leer discontinuities (HLLD) approximate Riemann
solver, while third order in time is achieved using a Runge-Kutta
scheme. \cite{2005JCoPh.208..315M} have shown that the HLLD algorithm
is very efficient in solving discontinuities formed in the MHD
system; consequently the adopted scheme is particularly appropriate
to describe the shocks formed during the interaction of the remnant
with the surrounding inhomogeneous medium. A monotonized central
difference limiter (the least diffusive limiter available in \PLUTO)
for the primitive variables is used. The solenoidal constraint of
the magnetic field is controlled by adopting an hyperbolic/parabolic
divergence cleaning technique available in \PLUTO\
(\citealt{2002JCoPh.175..645D, 2010JCoPh.229.5896M}).  The optically
thin radiative losses are included in a fractional step formalism
\citep{2007ApJS..170..228M}; in such a way, the $2^{nd}$ time
accuracy is preserved as the advection and source steps are at least
of the $2^{nd}$ order accurate. The radiative losses $\Lambda$
values are computed at the temperature of interest using a table
lookup/interpolation method. The code was extended by additional
computational modules to calculate the deviations from
temperature-equilibration between electrons and ions (by including
the almost instantaneous heating of electrons at shock fronts up to
$kT ~ 0.3$~keV by lower hybrid waves - \citealt{2007ApJ...654L..69G} - and
the effects of Coulomb collisions for the calculation of ion and electron
temperatures in the post-shock plasma; see \citealt{2015ApJ...810..168O}
for further details) and the deviations from equilibrium of ionization of
the most abundant ions (through the computation of the maximum ionization
age in each cell of the spatial domain; \citealt{2015ApJ...810..168O}).

Following \cite{2015ApJ...810..168O}, we assumed that the initial
(about 24 hours after the SN event) density structure of the ejecta
was clumpy, as also suggested by both theoretical studies (e.g.
\citealt{2000ApJS..127..141N, 2006A&A...453..661K, 2008ARA&A..46..433W,
2010A&A...521A..38G, 2015A&A...577A..48W}) and spectropolarimetric
studies of SNe (e.g. \citealt{2003ApJ...591.1110W, 2004ApJ...604L..53W,
2010ApJ...720.1500H}). Therefore, after the 1D radial density
distribution of ejecta (from model SN-M17-E1.2-N8; see above) was
mapped into 3D, the small-scale clumping of material is modeled as
per-cell random density perturbations derived from a power-law
probability distribution (\citealt{2012ApJ...749..156O,
2015ApJ...810..168O}). In our simulations, we considered ejecta
clumps with the same initial size $2\times 10^{12}$~cm (corresponding
to $\approx 1$\% of the initial remnant radius), and a maximum
density perturbation $\nu_{\rm max} = 5$.

The CSM around SN\,1987A is modeled as in \cite{2015ApJ...810..168O}.
In particular, we considered a spherically symmetric wind in the
immediate surrounding of the SN event which is characterized by a
gas density proportional to $r^{-2}$ (where $r$ is the radial
distance from the center of explosion), a mass-loss rate of
$\dot{M}_{\rm w} = 10^{-7} M_\odot$~year$^{-1}$, and a wind velocity
$u_{\rm w} = 500$~km~s$^{-1}$ (see also \citealt{2007Sci...315.1103M});
the termination shock of the wind is located approximately at $r_{\rm
w} = 0.05$~pc. The circumstellar nebula was modeled assuming that
it consists of an extended ionized H\,II region and a dense
inhomogeneous equatorial ring\footnote{The two external rings lying
below and above the equatorial plane are located outside the spatial
domain of our simulations and they do not affect the emission.} (see
Fig.~1 in \citealt{2015ApJ...810..168O}) composed by a uniform
smooth component and high-density spherical clumps mostly located
in its inner portion (e.g. \citealt{1995ApJ...452L..45C,
2005ApJS..159...60S}). The H\,II region has density $n_{\rm HII}$
and its inner edge in the equatorial plane is at distance $r_{\rm
HII}$ from the center of explosion.  The uniform smooth component
of the ring has density $n_{\rm rg}$, radius $r_{\rm rg}$, and an
elliptical cross section with the major axis $w_{\rm rg}$ lying on
the equatorial plane and height $h_{\rm rg}$; the $N_{\rm cl}$
clumps have a diameter $w_{\rm cl}$ and their plasma density and
radial distance from SN\,1987A are randomly distributed around the
values $<n_{\rm cl}>$ and $<r_{\rm cl}>$ respectively. In this paper
we adopted the parameters of the CSM derived in \cite{2015ApJ...810..168O}
to reproduce the X-ray observations (namely the morphology, lightcurves
and spectra) of SN\,1987A during the first $\approx 30$ years of
evolution. Table~\ref{tab1} summarizes the parameters adopted.

\begin{table}
\caption{Adopted parameters of the CSM for the MHD model of SN\,1987A.}
\label{tab1}
\begin{center}
\begin{tabular}{lclc}
\hline
\hline
CSM component & Parameters     & Units &  Best-fit values  \\
\hline
BSG wind: & $\dot{M}_{\rm w}$  & ($M_\odot$~year$^{-1}$) &  $10^{-7}$     \\
  & $u_{\rm w}$    &  (km~s$^{-1}$)   & 500    \\
  & $r_{\rm w}$    &  (pc)            & 0.05   \\
\hline
H\,II region: & $n_{\rm HII}$  &  ($10^2$ cm$^{-3}$) &   0.9     \\
 & $r_{\rm HII}$  &  (pc)              & 0.08    \\
\hline
Equatorial ring: & $n_{\rm rg}$   &  ($10^3$ cm$^{-3}$) &  1  \\
 & $r_{\rm rg}$   &  (pc)            & 0.18    \\
 & $w_{\rm rg}$   &  ($10^{17}$ cm)  & $1.7$   \\
 & $h_{\rm rg}$   &  ($10^{16}$ cm)  & $3.5$   \\
\hline
Clumps: & $<n_{\rm cl}>$ &  ($10^4$ cm$^{-3}$) &  $2.5\pm 0.3$    
  \\
 & $<r_{\rm cl}>$ &  (pc)            & $0.155\pm 0.015$     \\
 & $w_{\rm cl}$   &  ($10^{16}$ cm)  & $1.7$           \\
 & $N_{\rm cl}$   &                  &   50         \\
\hline
\end{tabular}
\end{center}
\end{table}

The simulations include passive tracers to follow the evolution of
the different plasma components (ejecta, H\,II region, and ring
material) and to store information on the shocked plasma (time,
shock velocity, and shock position, i.e. Lagrangian coordinates,
when a cell of the mesh is shocked either by the forward or by the
reverse shock) required to synthesize the thermal X-ray and non-thermal
radio emission (see Sects.~\ref{sec:x-ray} and \ref{sec:radio}).
The continuity equations of the tracers are solved in addition to
our set of MHD equations. In the case of tracers associated with
the different plasma components, each material is initialized with
$C_{\rm i} = 1$, while $C_{\rm i} = 0$ elsewhere, where the index
``i'' refers to the ejecta (ej), the H\,II region (HII), and the
ring material (rg). All the other tracers are initialized to zero
everywhere.

The initial computational domain is a Cartesian box extending between
$-4\times 10^{14}$~cm and $4\times 10^{14}$~cm in the $x$, $y$,
and $z$ directions. The box is covered by a uniform grid of $1024^3$
zones, leading to a spatial resolution of $\approx 8\times 10^{11}$~cm
($\approx 2.6\times 10^{-7}$~pc). The SN explosion is assumed to sit
at the origin of the 3D Cartesian coordinate system $(x_0, y_0, z_0) =
(0, 0, 0)$. In order to follow the large physical scales spanned during
the remnant expansion, we followed the same mesh strategy as proposed by
\cite{2013ApJ...773..161O} in the modeling of core-collapse SN explosions
(see also \citealt{1997ApJ...486.1026N}). More
specifically, the computational domain was gradually extended as the
forward shock propagates and the physical values were remapped in the
new domains. When the forward shock is close to one of the boundaries
of the Cartesian box, the physical size of the computational domain is
extended by a factor of 1.2 in all directions, maintaining a uniform grid
of $1024^3$ zones\footnote{As a result, the spatial resolution gradually
decreases by a factor 1.2 during the consecutive re-mappings.}. All the
physical quantities in the extended region are set to the values of the
pre-SN CSM. As discussed by \cite{2013ApJ...773..161O}, this approach
is possible because the propagation of the forward shock is supersonic
and it does not introduce errors larger than 0.1\% after 40 re-mappings.
We found that 43 re-mappings were necessary to follow the interaction
of the blast wave with the CSM during 40 years of evolution. The final
domain extends between $-10^{18}$~cm and $10^{18}$~cm in the $x$,
$y$, and $z$ directions, leading to a spatial resolution of $\approx
2\times 10^{15}$~cm ($\approx 6.5\times 10^{-4}$~pc). This strategy
guaranteed to have more than 400 zones per remnant radius during the
whole evolution. All physical quantities were set to the values of the
pre-SN CSM at all boundaries.

\subsection{Pre-supernova ambient magnetic field}

Our simulations include the effect of an ambient magnetic field.
This field is not expected to influence the overall expansion and
evolution of the blast wave which is characterized by a high plasma
$\beta$ (defined as the ratio between thermal pressure and magnetic
pressure). Nevertheless, a magnetic field is required for the
synthesis of radio emission (see Sect.~\ref{sec:radio}) and it might
play a role locally in preserving inhomogeneities of the CSM (as
the equatorial ring and the clumps) from complete fragmentation (by
limiting the growth of HD instabilities; e.g.
\citealt{2008ApJ...678..274O}). For these reasons, we introduced
a seed magnetic field in our simulations.

The characteristics of magnetic fields around massive stars, and the
way these fields, in combination with stellar rotation, confine stellar
wind outflows of massive stars (thus building up an extended circumstellar
magnetosphere) have been studied in the literature through accurate MHD
simulations (e.g. \citealt{2005MNRAS.357..251T, 2005ApJ...630L..81T,
2008MNRAS.385...97U, 2013MNRAS.428.2723U}). However, at present, there
is no hint about the initial magnetic field strength and configuration
around Sk $−69^{\rm o} 202$ which resulted from the merging of two
massive stars. Therefore we assumed here the simplest and most
common magnetic field configuration around a rotating star.

Most likely, the ambient magnetic field is that originating from
the stellar progenitor. \cite{2015A&A...584A..54P} have
argued that BSGs descend from magnetic massive stars. These stars
can result from strong binary interaction and, in particular, from
the merging of two main sequence stars (\citealt{2012Sci...337..444S}),
as it is the case for Sk $−69^{\rm o} 202$ (e.g.
\citealt{2007Sci...315.1103M}). In magnetic massive stars, the
observed fields are, in general, large scale fields with a dominant
dipole component and with typical values at the stellar surface
varying in the range between 100 G and 10 kG (see the review by
\citealt{2009ARA&A..47..333D}). Due to the rotation of the progenitor
and to the expanding stellar wind, the field is expected to be
twisted around the rotation axis of the progenitor, even if magnetic
massive stars generally rotate more slowly than normal stars of the
same mass, by factors of ten to 1000 (e.g. \citealt{2009ARA&A..47..333D}).
Thus we expect that the field might be characterized by a toroidal
component resulting in the so-called ``Parker spiral'', in analogy
with the spiral-shaped magnetic field on the interplanetary medium
of the solar system (\citealt{1958ApJ...128..664P}). For the
purposes of the present paper, we assumed as initial ambient magnetic
field, a Parker spiral which, in spherical coordinates $(r, \theta,
\phi)$, can be described by the radial and toroidal components:

\begin{equation}
B_{\rm r} = \frac{A_1}{r^2}~, ~~~~~~~~
B_{\phi} = -\frac{A_2}{r} \sin\theta
\label{b_spiral}
\end{equation}

\noindent
\noindent
where

\begin{equation}
A_1 = B_0 r_0^2~, ~~~~~~~~ A_2 = B_0 r_0^2\omega_{\rm s}/u_{\rm w}~,
\label{wind_const}
\end{equation}

\noindent
$B_0$ is the average magnetic field strength at the stellar surface,
$r_0$ the stellar radius, $\omega_{\rm s}$ the angular velocity of
stellar rotation, and $u_{\rm w}$ the wind speed. In the case of
SN\,1987A, the field configuration probably resulted from the
different phases in the evolution of the stellar progenitor whose
details are poorly known; thus the field configuration can be more
complex than that described here. Recently, \cite{2018ApJ...861L...9Z}
have estimated the strength of the ambient (pre-shock) magnetic
field within the nebula $\approx 30\, \mu$G and the (post-shock)
field strength within the remnant of the order of a few mG. In the
light of this, we considered two cases in which the field strength
is almost the same in proximity of the initial remnant ($t\approx
24$~hours), whereas it is either of the order of $1\,\mu$G (run
MOD-B1) or of the order of $100\,\mu$G (run MOD-B100) at the inner
edge of the nebula (at a distance from the center of explosion
$r\approx 0.08$~pc). This is realized by considering the parameter
$A_1 = 3\times 10^{28}$~G~cm$^{2}$ in both cases, and $A_2 = 8\times
10^{10}$~G~cm in MOD-B1 and $A_2 = 8\times 10^{13}$~G~cm in MOD-B100.
If $r_0 \approx 45\,R_{\odot}$ is the radius of the BSG progenitor
of SN\,1987A (\citealt{1988ApJ...330..218W}), from Eq.~\ref{wind_const}
$B_0\approx 3$~kG in both models, a value well within the range
inferred for magnetic massive stars (e.g. \citealt{2009ARA&A..47..333D}).

As expected, the adopted field strength does not influence
significantly the evolution of the remnant, being the plasma $\beta >
10^5$ at the forward shock after the breakout of the shock wave at the
stellar surface. By comparing the SNR ram pressure, thermal pressure,
and ambient magnetic field pressure soon after the shock breakout,
we note that the remnant expansion and dynamics could have been
significantly affected by the ambient magnetic field if its strength
close to the stellar surface had been larger than 1~MG, a value much
higher than the typical field strengths observed in massive stars
(e.g. \citealt{2009ARA&A..47..333D}).

Finally, it is worth noting that our Eqs.~\ref{b_spiral}-\ref{wind_const}
describe the magnetic field also in the initial ($t\approx 24$~hours)
remnant interior. We expect, however, that a more realistic field there
(especially in the immediate surroundings of the remnant compact
object) would be much more complex than that adopted here and it
should reflect the field of the stellar interior before the collapse
of the stellar progenitor. Current stellar evolution models
predict that differentially rotating massive stars have relatively
weak fields before collapse (e.g. \citealt{2005ApJ...626..350H}).
Nevertheless several mechanisms for efficient field amplification after
core bounce may be present (e.g. \citealt{2014MNRAS.445.3169O,
2018PhRvD..98h3018M}, for non-rotating progenitors;
\citealt{2017MNRAS.469L..43O}, for rotating models). Among these,
magneto-rotational instability (MRI) driven by the combined
action of magnetic field and rotation may produce dynamically
relevant fields after bounce which can affect the evolution of
the SN (e.g. \citealt{2012ApJ...759..110M, 2016MNRAS.460.3316R,
2016MNRAS.456.3782R}).  In the case of a star with mass similar to
that presumed for Sk $−69^{\rm o} 202$ ($\approx 20 M_{\odot}$),
\cite{2018JPhG...45h4001O} have explored the effects of magnetic field and
rotation on the core collapse by considering different, artificially added
profiles of rotation and magnetic field. They found that the evolution
of the shock wave can be modified if the field is in equipartition with
the gas pressure: the explosion geometry is bipolar with two outflows
propagating along the rotational axis and downflows at low latitudes. The
strength of the magnetic field (in combination with rotation) therefore
can have significant effects on the evolution of the SN. In the case
of SN 1987A, however, we do not have any indication about the strength
and configuration of the stellar magnetic field during the core-collapse
and the expansion of the shock wave through the stellar interior. Thus,
for the sake of simplicity, we assumed here that the field is weak
enough that the magnetic energy density is much lower than the kinetic
energy of ejecta or, in other words, that the field has no significant
influence on the ejecta dynamics.

\begin{figure}[!ht]
  \begin{center}
    \leavevmode
        \epsfig{file=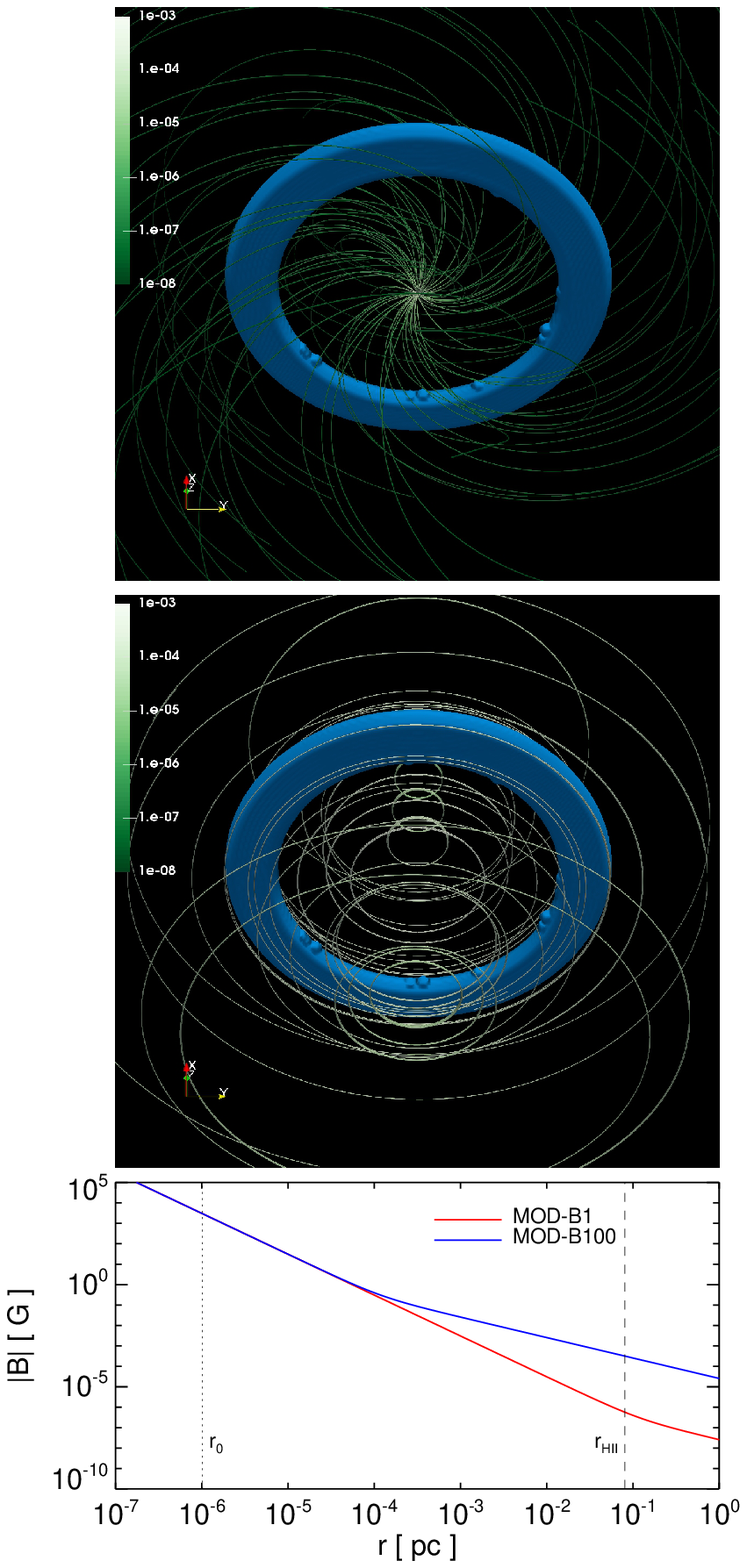, width=8.cm}
	\caption{Initial configuration of the pre-SN ambient magnetic
	field for models MOD-B1 (upper panel) and MOD-B100 (middle
	panel). The green lines are sampled magnetic field lines; 
	the scale of the magnetic field strength is log and is given
	by the bar on the left of each panel, in units of $G$.
	The unshocked equatorial ring material (with $n \geq
	1000$~cm$^{-3}$) is marked blue; the figure does not show
	the H\,II region around the ring. The lower panel shows the
	radial magnetic field strength in the equatorial plane for
	the two models; the dotted line marks the radius of the BSG
	progenitor, the dashed line marks the inner edge of the
	H\,II region.}
  \label{fig1}
\end{center}
\end{figure}

Figure~\ref{fig1} shows the initial configuration of the ambient
magnetic field in models MOD-B1 and MOD-B100. In the latter model,
the toroidal component of the magnetic field is largely dominant.

\subsection{Synthesis of thermal X-ray emission}
\label{sec:x-ray}

We followed the approach outlined in \cite{2015ApJ...810..168O} to
synthesize, from the model results, the thermal X-ray emission
originating from the impact of the blast wave with the nebula. In
this section, we summarize the main steps in this approach; we
refer the reader to Sect.~2.3 in \cite{2015ApJ...810..168O} for
more details (see also \citealt{orlando2, 2009A&A...493.1049O}).

As a first step, we rotated the system about the three axis to fit
the orientation of the ring with respect to the line-of-sight (LoS)
as found from the analysis of optical data (\citealt{2005ApJS..159...60S}):
$i_x = 41^{\rm o}$, $i_y = -8^{\rm o}$, and $i_z = -9^{\rm o}$.
Then, for each $j$th cell of the spatial domain, we derived: a) the
emission measure as em$_{\rm j} = n_{\rm Hj}^2 V_{\rm j}$ (where
$n_{\rm Hj}^2$ is the hydrogen number density in the cell, $V_{\rm
j}$ is the cell volume, and we assume fully ionized plasma); b) the
maximum ionization age as $\tau_{\rm j} = n_{\rm ej} \Delta t_{\rm
j}$ (where $\Delta t_{\rm j}$ is the time since the plasma in the
$j$-th domain cell was shocked); c) the electron temperature $T_{\rm
ej}$ from the ion temperature, plasma density, and $\Delta t_{\rm
j}$, by assuming Coulomb collisions and starting from an electron
temperature at the shock front $kT = 0.3$~keV which is assumed to
be the same at any time as a result of instantaneous heating by
lower hybrid waves (\citealt{2007ApJ...654L..69G}; see also
\citealt{2015ApJ...810..168O}). In calculations
of the electron heating and ionization time-scale, the forward and
reverse shocks are treated in the same way. From the values of
emission measure, electron temperature, and maximum ionization age
in the $j$th domain cell, we synthesized the X-ray emission in the
$[0.1, 10]$~keV band by using the NEI (non-equilibrium of ionization)
emission model VPSHOCK available in the XSPEC package along with the
NEI atomic data from ATOMDB (\citealt{2001ApJ...556L..91S}).

We assumed the source at a distance $D = 51.4$ kpc
(\citealt{1999IAUS..190..549P}) and adopted the metal abundances
derived by \cite{2009ApJ...692.1190Z} from the analysis of deep
{\it Chandra} observations of SN\,1987A. We filtered the X-ray
spectrum from each cell through the photoelectric absorption by the
ISM, assuming a column density $N_{\rm H} = 2.35\times 10^{21}$
cm$^{-2}$ (\citealt{2006ApJ...646.1001P}). Finally, we integrated
the absorbed X-ray spectra from the cells in the whole spatial
domain and folded the resulting integrated spectrum through the
instrumental response of either {\it XMM-Newton}/EPIC or {\it
Chandra}/ACIS, obtaining the relevant focal-plane spectra. Since
the synthetic data are put in a format virtually identical to
that of true X-ray observations, we analyzed the synthetic observations
with the standard data analysis system used for {\it XMM-Newton}
and {\it Chandra}.

\subsection{Synthesis of non-thermal radio emission}
\label{sec:radio}

We synthesized the radio emission arising from the interaction of
the blast wave with the nebula by using REMLIGHT, a code for the
synthesis of synchrotron radio, X-ray, and inverse Compton $\gamma$-ray
emission from MHD simulations, in the general case of a remnant
expanding through an inhomogeneous ambient medium and/or a non-uniform
ambient magnetic field (\citealt{2007A&A...470..927O,
2011A&A...526A.129O}). Since REMLIGHT does not take into account
the synchrotron self-absorption (SSA) and the free-free absorption
(FFA), we evaluated the importance of these effects in the radio
emission by using the CR-hydro-NEI code (\citealt{2012ApJ...750..156L})
and the SPEV (SPectral EVolution) code (\citealt{2009ApJ...696.1142M,
2015HEDP...17...92O}). CR-hydro-NEI is a 1D code which includes,
among other things: a momentum- and space-dependent CR diffusion
coefficient, the magnetic field amplification, the deviations from
equilibrium of ionization, the SSA and FFA. SPEV is a 3D parallel
code to solve the non-thermal electron transport and the evolution
equations which includes: the time- and frequency-dependent radiative
transfer in a dynamically changing background, and the effects of SSA.

The radio emissivity in REMLIGHT is expressed as (e.g.
\citealt{1965ARA&A...3..297G})

\begin{equation}
i(\nu)=C_r K B_{\bot}^{\,\alpha+1}\nu^{\,-\alpha},
\label{radio}
\end{equation}

\noindent
where $C_r$ is a constant, $K$ is the normalization of the electron
distribution, $B_{\bot}$ is the component of the magnetic field
perpendicular to the LoS, $\nu$ is the frequency of the radiation,
and $\alpha$ is the synchrotron spectral index. The code describes
the post-shock evolution of relativistic electrons by adopting the
model of \cite{1998ApJ...493..375R}; with this approach, $K$ can
be expressed as (Eq.~A.8 in \citealt{2007A&A...470..927O}):

\begin{equation}
 \frac{K(a,t)}{K\rs{s}(R,t)}=
 \left(\frac{\rho\rs{o}(a)}{\rho\rs{o}(R)}\right)
 \left(\frac{V\rs{sh}(t)}{V\rs{sh}(t\rs{i})}\right)^{b}
 \left(\frac{\rho(a,t)}{\rho\rs{s}(a,t\rs{i})}\right)^{2\alpha/3+1}.
 \label{KK}
\end{equation}

\noindent
where

\begin{equation}
K\rs{s}(R,t) = C\rs{norm} \rho\rs{s}(R) V\rs{sh}(t)^{-b},
\label{KK_norm}
\end{equation}

\noindent
$C\rs{norm}$ is a parameter, $a\equiv R(t_i)$ is the Lagrangian
coordinate, $R$ is the shock radius, $t$ is the current time, $t_i$
is the time when the fluid element was shocked, $\rho$ is the gas
density, $V_{\rm sh}$ is the shock velocity, $b$ regulates the
dependence of $K$ on the shock velocity (\citealt{1998ApJ...493..375R}),
and the labels ``s'' and ``o'' refer to the immediately post-shock
and pre-shock values, respectively. By substituting Eqs. \ref{KK}
and \ref{KK_norm} in Eq. \ref{radio}, we obtain

\begin{equation}
i(\nu)=C_r C\rs{norm} \sigma^{-2\alpha/3} \rho\rs{o}(a)
V\rs{sh}(t\rs{i})^{-b}
\left(\frac{\rho(a,t)}{\rho\rs{o}(a)}\right)^{2\alpha/3+1}
B_{\bot}^{\,\alpha+1}\nu^{\,-\alpha}
\label{eq5}
\end{equation}

\noindent
where $\sigma = (\gamma+1)/(\gamma-1)$. Finally, we rewrite Eq.
\ref{eq5} as:

\begin{equation}
i(\nu)= C'(\nu) K' B_{\bot}^{\,\alpha+1}
\label{eq6}
\end{equation}

\noindent
where

\begin{equation}
C'(\nu) = C_r C\rs{norm} \sigma^{-2\alpha/3} \nu^{\,-\alpha}
\label{c_prime} 
\end{equation}

\begin{equation}
K' = \rho\rs{o}(a) V\rs{sh}(t\rs{i})^{-b}
\left(\frac{\rho(a,t)}{\rho\rs{o}(a)}\right)^{2\alpha/3+1}
\end{equation}

\noindent
in order to include all variables not depending on the model in
$C'$. The values of $a$, $t_i$, and $V_{\rm sh}$ are stored in
passive tracers included in the calculations. The evolution of
electron spectra from the forward and reverse shocks are described
by the same Eq. \ref{eq6} where all values are taken for the
respective shock. In principle, the parameter $C'$ can be different
for forward and reverse shocks: $C'\rs{FS}$ and $C'\rs{RS}$. The
latter two free parameters balance the contributions to the overall
radio emission from electron populations accelerated by each shock.

Then the total radio intensity (Stokes parameter $I$) at a given
frequency $\nu_0$ is derived by integrating the emissivity $i(\nu_0)$
along the LoS:

\begin{equation}
I(\nu_0) = \int i(\nu_0) ~dl~,
\end{equation}

\noindent
where $dl$ is the increment along the LoS.

ATCA observations have shown that the radio spectral index in SN\,1987A
is not constant and evolves between day 1517 and 8014 after explosion,
ranging between $\approx 1$ and 0.7 (\citealt{2010ApJ...710.1515Z}). We
synthesized the radio emission either considering $\alpha$ constant
($\alpha = 0.9$ or 0.7) or including its evolution by adopting the
spectral index fit calculated by \cite{2010ApJ...710.1515Z} from day
2511 as:

\begin{equation}
\alpha(t) = \alpha_0 + \beta_0 \times \frac{t - t_0}{\Delta t}~,
\label{eq:alpha}
\end{equation}

\noindent
where $\alpha_0 = 0.825$, $\beta_0 = -0.018$, $t$ is expressed in
days, $t_0 = 5000$ days, and $\Delta t = 365$ days; in addition we
assumed $\alpha(t) = 0.95$ before day 2511, and $\alpha(t) = 0.5$
after day 11590.

The parameter $b$ in Eq.~\ref{KK} is a constant and determines how
the injection efficiency (the fraction of electrons that move into
the CR pool) depends on the shock properties.
\cite{1998ApJ...493..375R} considered values of $b$ ranging between
0 and $-2$, thus assuming that the injection efficiency behaves in
a way similar to acceleration efficiency: stronger shocks might
inject particles more effectively. \cite{2011MNRAS.413.1657P} showed
that the smaller $b$, the thicker the radial profiles of the surface
brightness in all bands, an effect that is most prominent in the
radio band. Here we explored two cases, $b = 0, -1$. Finally, we
assumed no dependence of the particle injection on the obliquity
angle $\Theta$ (i.e., the angle between the unperturbed ambient
magnetic field and the shock normal), namely the isotropic scenario.
This particular choice of the obliquity dependence is almost
unimportant for the purposes of the present paper because we
considered the radio fluxes from the whole SNR (in this case the spatial
features from obliquity are absorbed in the coefficient $C_{\rm norm}$).

\begin{figure*}[!ht]
  \begin{center}
    \leavevmode
        \epsfig{file=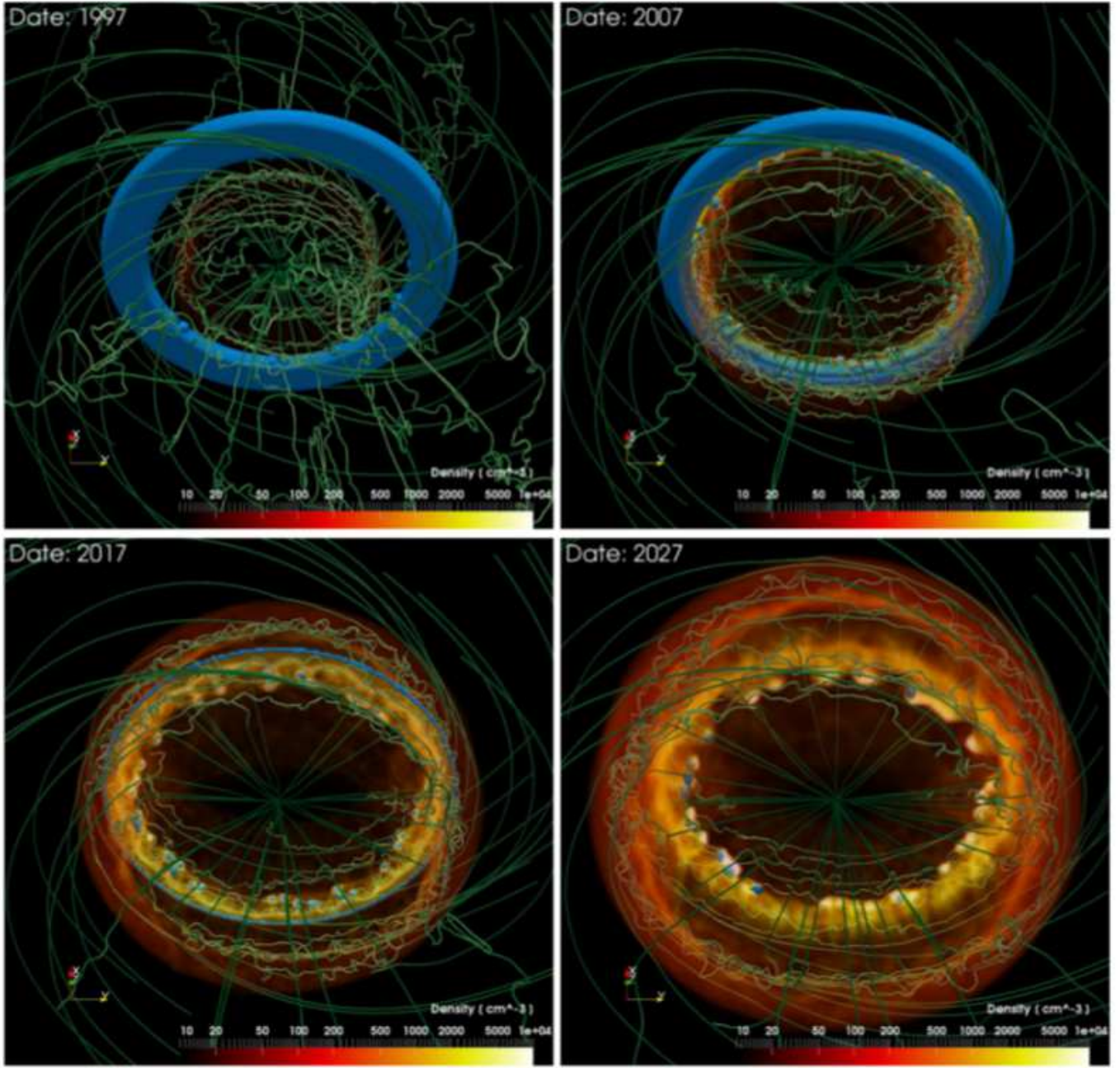, width=18.cm}
	\caption{Three-dimensional volume rendering of the number
	density during the interaction of the blast wave with the
	magnetized nebula at the labeled times for model MOD-B1.
	The density of the shocked plasma is marked red and that
	of the unshocked equatorial ring material (with $n \geq
	1000$~cm$^{-3}$) is marked blue. The green lines are sampled
	magnetic field lines; the scale of the magnetic field strength
	is log and is the same of Fig.~\ref{fig1}. Refer to on-line
	Movie 1 for an animation of these data.}
  \label{fig_maps}
\end{center} \end{figure*}

\section{Results}
\label{sec:result}

As mentioned in Sect.~\ref{sec:model}, the first 24 hours in the
post-explosion evolution of the SN are described by the best-fit
model (run SN-M17-E1.2-N8) presented in \cite{2015ApJ...810..168O}.
This model reproduces the main observable of the SN (i.e., bolometric
lightcurve, evolution of line velocities, and continuum temperature
at the photosphere).  We refer the reader to \cite{2015ApJ...810..168O}
for a complete description of the evolution in this phase. Then we
followed the transition from the SN to the SNR phase during the
subsequent 40 years, including an ambient magnetic field. From the
evolution of the remnant and its interaction with the nebula, we
did not find any appreciable difference between the two cases
considered, namely MOD-B1 and MOD-B100. In the following, therefore,
we discuss in detail the results of run MOD-B1, mentioning the
differences (if any) with the other case.

Figure~\ref{fig_maps} shows the evolution of number density during
the interaction of the blast wave with the magnetized nebula for
model MOD-B1. A movie showing the 3D rendering of plasma density
(in units of cm$^{-3}$) distribution during the blast evolution is
provided as on-line material (Movie 1). We found that the evolution
in the present model (which includes the magnetic field) follows
the same trend described in detail by the HD model of
\cite{2015ApJ...810..168O} and can be characterized by three main
phases: an H\,II-region-dominated phase in which the fastest ejecta
interact with the H\,II region (see upper left panel in
Fig.~\ref{fig_maps}; the unshocked H\,II region is not included in
the rendering); a ring-dominated phase, in which the dynamics is
dominated by the interaction of the blast wave with the dense
equatorial ring (upper right and lower left panels; the unshocked
ring is marked blue); and an ejecta-dominated phase, in which the
forward shock propagates beyond the majority of the dense ring
material and the reverse shock travels through the inner envelope
of the SN (lower right panel).

\subsection{Effects of the ambient magnetic field}
\label{sec:field}

As expected, the presence of an ambient magnetic field (of the
assumed strength) does not change significantly the overall evolution
of the remnant. In fact, for the values of explosion energy and
ambient field strength considered, the kinetic energy of the shock
is orders of magnitude larger than the energy density in the ambient
$B$ field (even in MOD-B100). As a result, the magnetic field in
the remnant interior is stretched outward by the expanding blast
wave and assumes an almost radial configuration (see Fig.~\ref{fig_maps}),
at variance with the toroidal field in the pre-shock nebula (the Parker
spiral; Fig.~\ref{fig1}).

\begin{figure}
  \begin{center}
    \leavevmode
        \epsfig{file=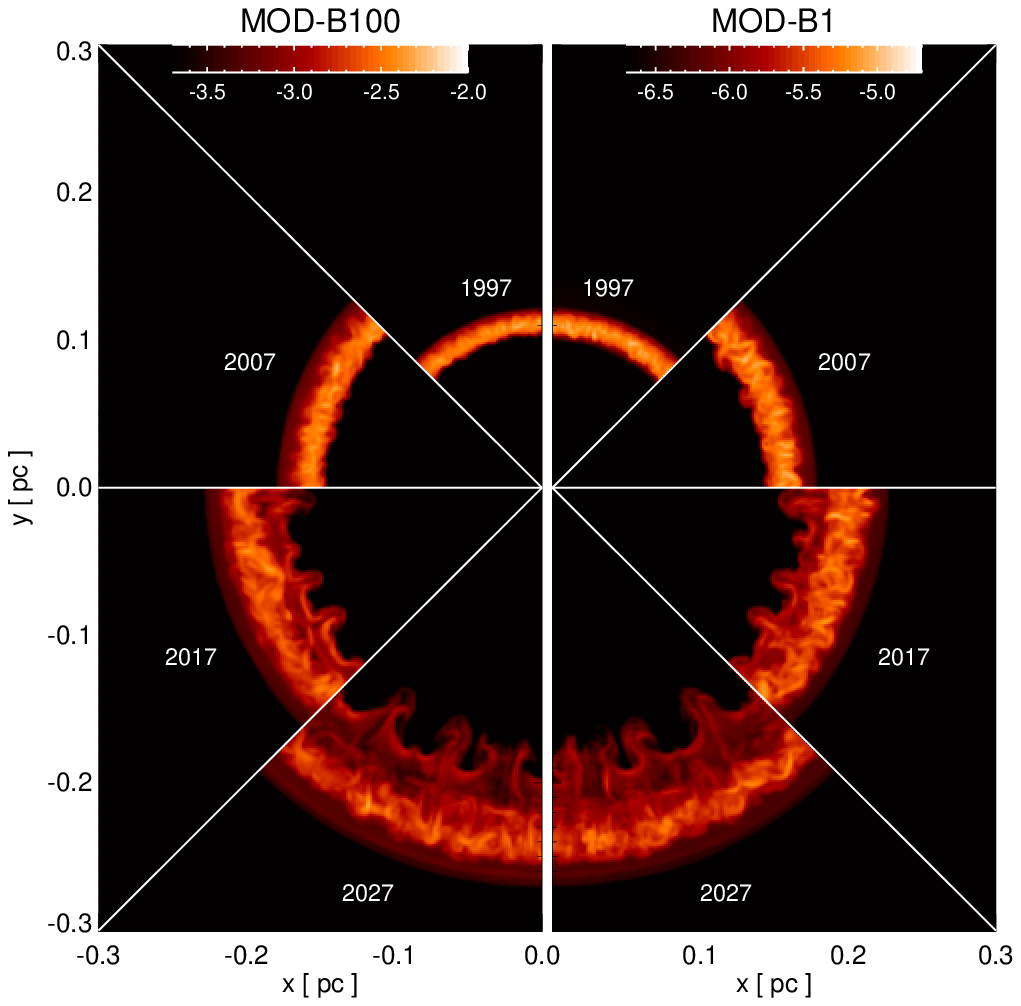, width=9cm}
        \epsfig{file=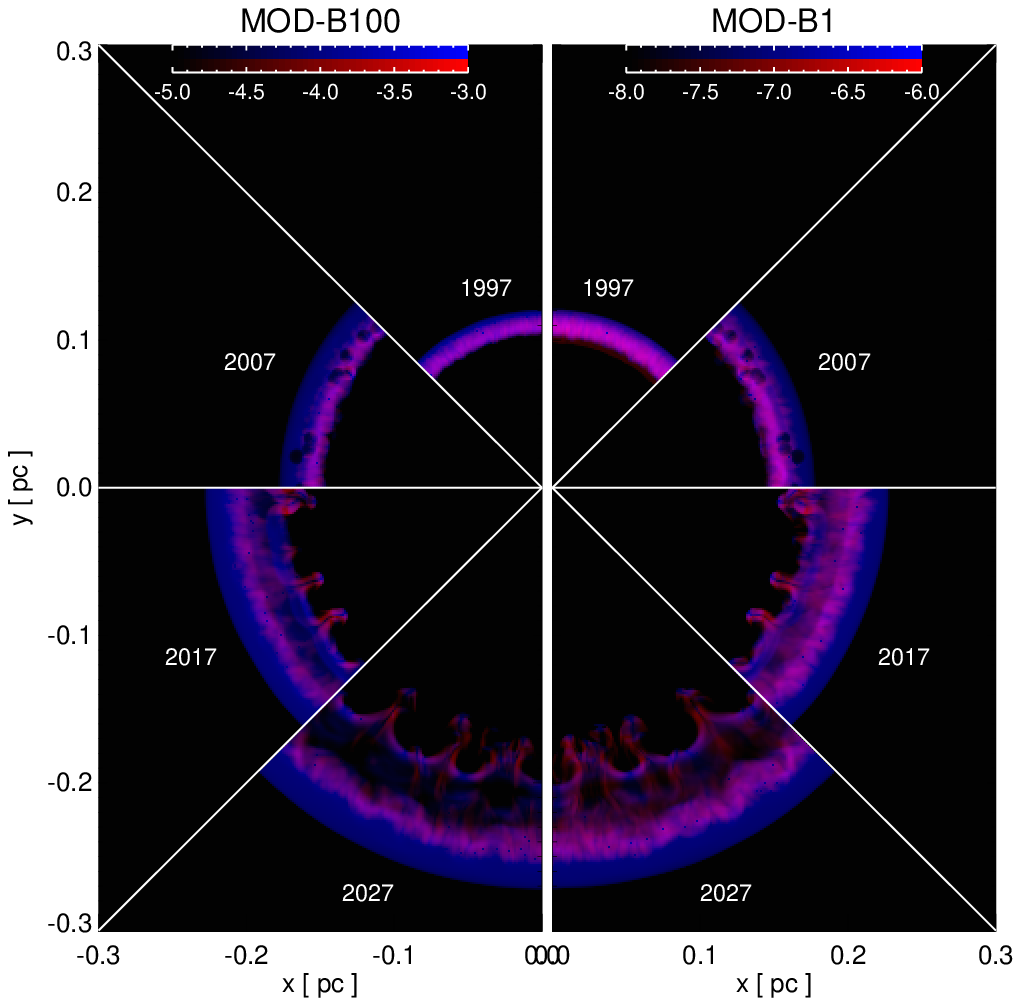, width=9cm}
	\caption{Upper panel: maps in the $(x,y)$ plane of maximum
	magnetic field strength along $z$ at the labeled times for
	MOD-B100 (on the left) and MOD-B1 (on the right). Lower
	panel: as in the upper panel, but for a two-color composite
	image showing the density averaged radial (red) and toroidal
	(blue) components of the magnetic field along $z$ (only
	shocked cells have been considered). The scale of the
	magnetic field strength is log and is given by the bar on
	the top of each panel, in units of $G$.}
  \label{fig_bvol}
\end{center} \end{figure}
\begin{figure}
  \begin{center}
    \leavevmode
        \epsfig{file=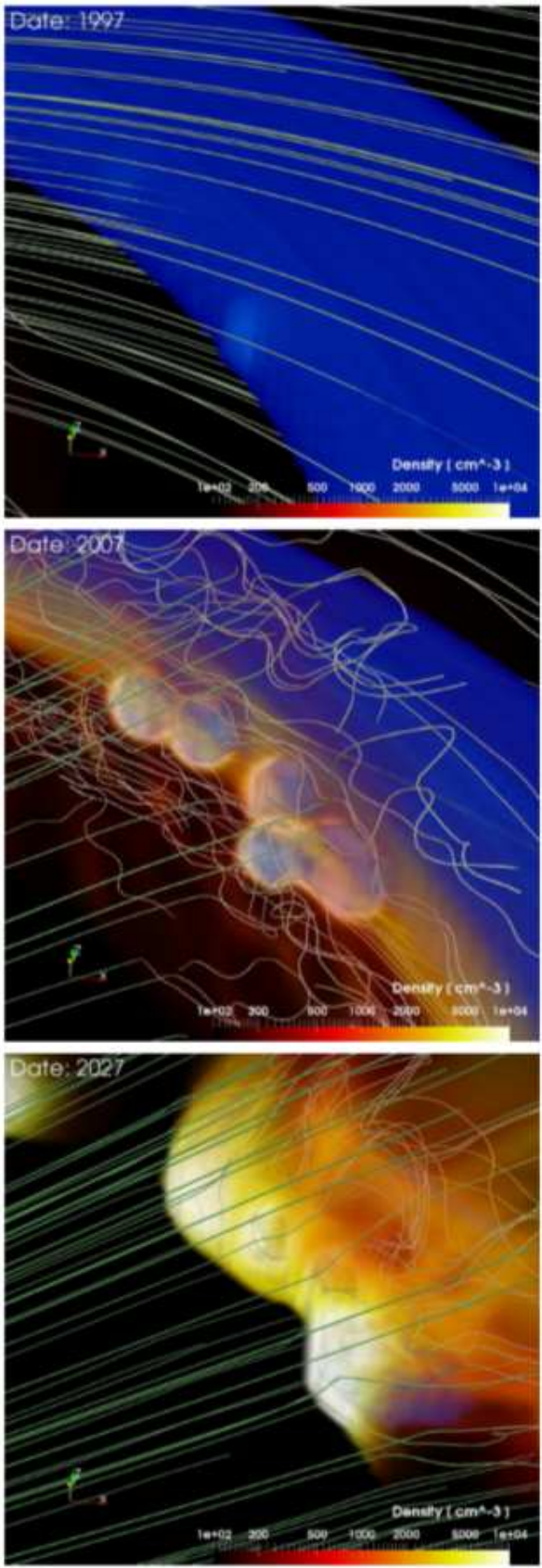, width=7.8cm}
	\caption{Same as in Fig.~\ref{fig_maps} for a close-up view
	of the interaction of the blast wave with the ring. Refer
	to on-line Movie 2 for an animation of these data.}
  \label{fig_zoom}
\end{center}
\end{figure}

\begin{figure*}[!ht]
  \begin{center}
    \leavevmode
        \epsfig{file=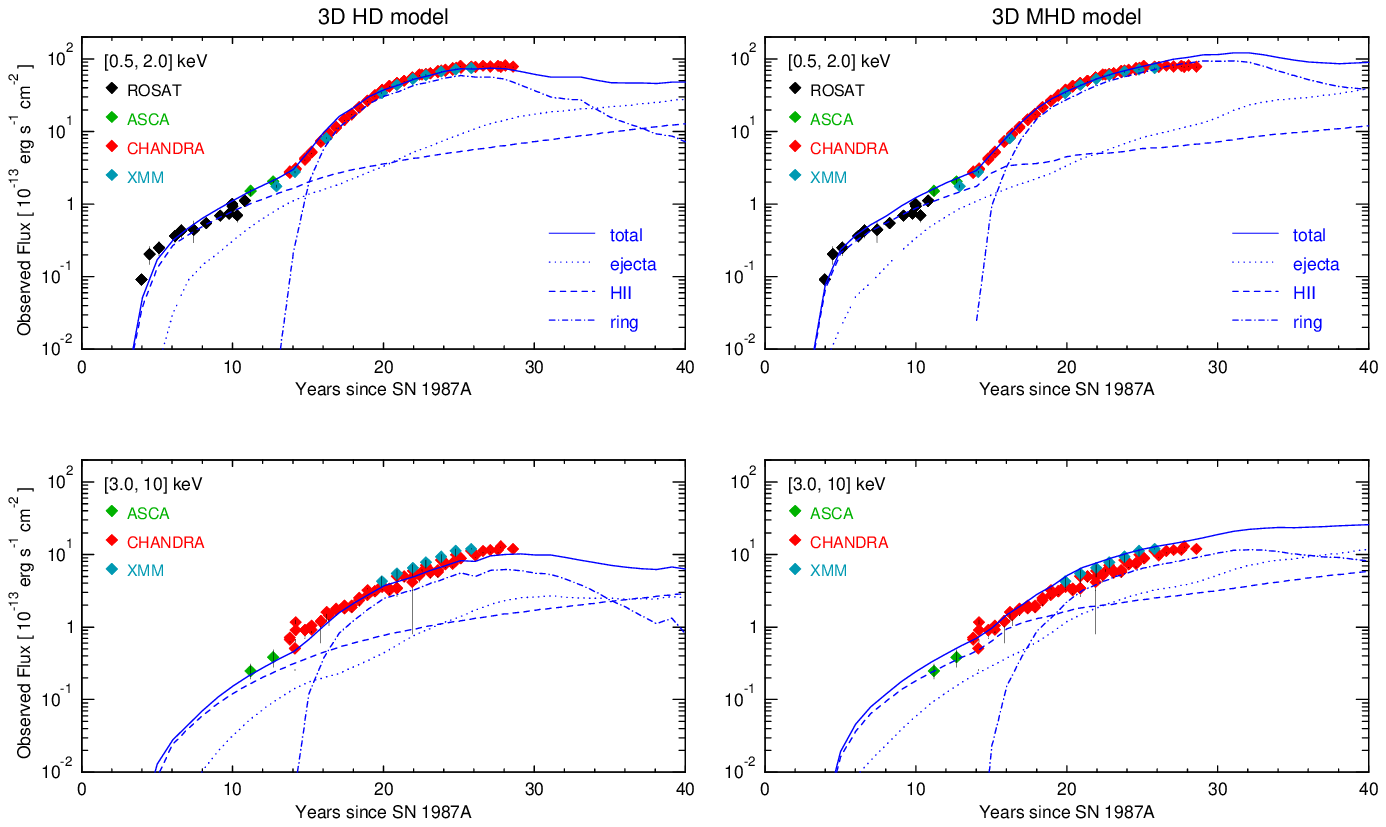, width=18.cm}
	\caption{Comparison between observed (symbols) and synthetic
	(lines) X-ray lightcurves in the $[0.5, 2]$~keV (upper
	panels) and $[3, 10]$~keV (lower panels) bands for the
	best-fit HD model of \cite{2015ApJ...810..168O}
	(left panels) and for the MHD model MOD-B1 presented here (right
	panels). Solid lines show the synthetic lightcurves; dotted,
	dashed and dot-dashed lines mark the contribution to emission from
	the shocked ejecta, the shocked plasma from the H\,II region,
	and the shocked plasma from the ring, respectively; the
	different symbols show the fluxes measured with Rosat (black
	diamonds; \citealt{2006A&A...460..811H}), {\it ASCA} (green;
	\citealt{2015ApJ...810..168O}), {\it Chandra} (red;
	\citealt{2013ApJ...764...11H, 2016ApJ...829...40F}) and
	{\it XMM-Newton} (cyan; \citealt{2006A&A...460..811H,
	2012A&A...548L...3M, 2015ApJ...810..168O}).}
  \label{fig_lc}
\end{center} \end{figure*}

In the mixing region between the forward and reverse shocks, the
magnetic field follows the plasma structures formed during the
growth of Rayleigh-Taylor (RT) instabilities at the contact
discontinuity and shows a turbulent structure (see Fig.~\ref{fig_maps}).
The latter structure is enhanced even more by the expanding small-scale
clumps of ejecta which modify the $B$ field in the mixing region
(e.g. \citealt{2012ApJ...749..156O}). The upper panel in
Fig.~\ref{fig_bvol} shows the maximum magnetic field strength along
the $z$-axis for both MOD-B1 and MOD-B100; the lower panel shows
the density-averaged radial (red) and toroidal (blue) post-shock
components of $B$ along $z$. During the interaction of the blast
wave with the nebula, the magnetic field strength can be locally
enhanced by more than one order of magnitude, up to values of the
order of $|B|_{\rm max}\approx 30\,\mu$G in MOD-B1 and $|B|_{\rm
max}\approx 10\,$mG in MOD-B100, even if the back-reaction of
accelerated CRs is neglected in our calculations (see upper
panel in Fig.~\ref{fig_bvol}). This is mainly due to high compression
of magnetic field lines in the post-shock. In addition, a significant
radial component of $B$ appears in the mixing region, although the
pre-shock nebula is largely dominated by the toroidal component (see
lower panel in Fig.~\ref{fig_bvol}). This is due,
on one hand, to RT instabilities developing at the contact discontinuity
which lead to a preferentially radial component of $B$ around the
RT fingers (e.g. \citealt{2012ApJ...749..156O}) and, on the other
hand, to stretching of the magnetic field trapped at the border of
the ring and of the dense clumps by the plasma flow in the H\,II
region. Interestingly, a recent detection of linear polarization
of the synchrotron radio emission of SN\,1987A suggests a primarily
radial magnetic field across the remnant (\citealt{2018ApJ...861L...9Z}).
Such a radial magnetic field seems to be a common feature in young
SNRs (\citealt{1976AuJPh..29..435D, 2015A&ARv..23....3D}).

The magnetic field configuration is strongly modified
during the interaction of the blast wave with the dense equatorial
ring.  Fig.~\ref{fig_zoom} shows a close-up view of the interaction
(the on-line Movie 2 shows the 3D rendering of plasma density, in
units of cm$^{-3}$, during the blast evolution). As mentioned above,
after the impact of the blast wave onto the ring, the magnetic field
is trapped at the inner edge of the ring, leading to a continuous
increase of the magnetic field tension there. Even if the plasma
$\beta$ is, on average, larger than one\footnote{Close to the reverse
shock, the plasma $\beta$ can reach values lower than one.} (thus
not affecting the overall evolution of the remnant), the field
tension maintains a more laminar flow around the ring (and the
denser clumps of the ring), limiting the growth of HD instabilities
that would develop at the ring (and clumps) boundaries (see also,
\citealt{1994ApJ...433..757M, 1996ApJ...473..365J, 2005ApJ...619..327F,
2008ApJ...678..274O}). As a result, the magnetic field largely
limits the progressive erosion and fragmentation of the ring and
the latter survives for a longer time than if the magnetic field
was not present (as it happens in the HD model of
\citealt{2015ApJ...810..168O}).

Since the thermal X-ray emission (especially in the soft band) is
largely dominated by the shocked ring material during the ring-dominated
phase (e.g. \citealt{2015ApJ...810..168O}), we expect some changes
in the X-ray lightcurves due to the presence of the magnetic field.
Fig.~\ref{fig_lc} compares the X-ray lightcurves derived with our
3D MHD model\footnote{We found that MOD-B100 produces thermal X-ray lightcurves
almost identical to those produced by run MOD-B1.} (run MOD-B1)
with those derived with the 3D HD model of \cite{2015ApJ...810..168O};
the symbols mark the observed lightcurves. In both cases, the
transition from SN to SNR enters into the first phase of evolution
(H\,II-region-dominated phase) about three years after explosion,
when the ejecta reach the H\,II region and the X-ray flux rise
rapidly. During this phase the two models (HD and MHD) produce very
similar results and the effects of the magnetic field can be
considered to be negligible: the models reproduce quite well the
observed fluxes and slope of the lightcurves and the emission is
dominated by shocked plasma from the H\,II region and by a smaller
contribution from the outermost ejecta (dashed and dotted lines,
respectively, in Fig.~\ref{fig_lc}). This phase ends after $\approx
14$~years when the blast hits the first dense clump of the equatorial
ring.

Around year 14 the SNR enters into the second phase of evolution
(ring-dominated phase). At the beginning of the interaction with
the ring (between year 14 and $\approx 25$), the two models show
quite similar results, reproducing the observed lightcurves quite
well, although the MHD model predicts slightly higher X-ray fluxes
in the hard band. These higher fluxes are mainly due to higher
fluxes from the shocked ring (compare dot-dashed lines in the lower
panels of Fig.~\ref{fig_lc}). After year 25, the differences in the
fluxes predicted with the two models slightly increase with time:
both soft and hard X-ray fluxes are higher in the MHD model than
in the HD model. Again the differences originate mainly from the
contribution to X-ray emission from the shocked ring. In the HD
model, this contribution reaches a maximum between years 25 and 30;
then, it decreases rapidly in both bands, and it is no longer the
dominant component after year 34 when the SNR enters in the
ejecta-dominated phase (see left panels of Fig.~\ref{fig_lc}).
Instead, in the MHD model, the contribution from the ring decreases
slowly after the maximum around year 30 and it dominates the X-ray
emission until the end of the simulation (year 40) when it is
comparable to the contribution from the shocked ejecta (see right
panels of Fig.~\ref{fig_lc}). In the MHD model, the SNR enters in
the ejecta dominated phase after year 40.

\begin{figure}
  \begin{center}
    \leavevmode
        \epsfig{file=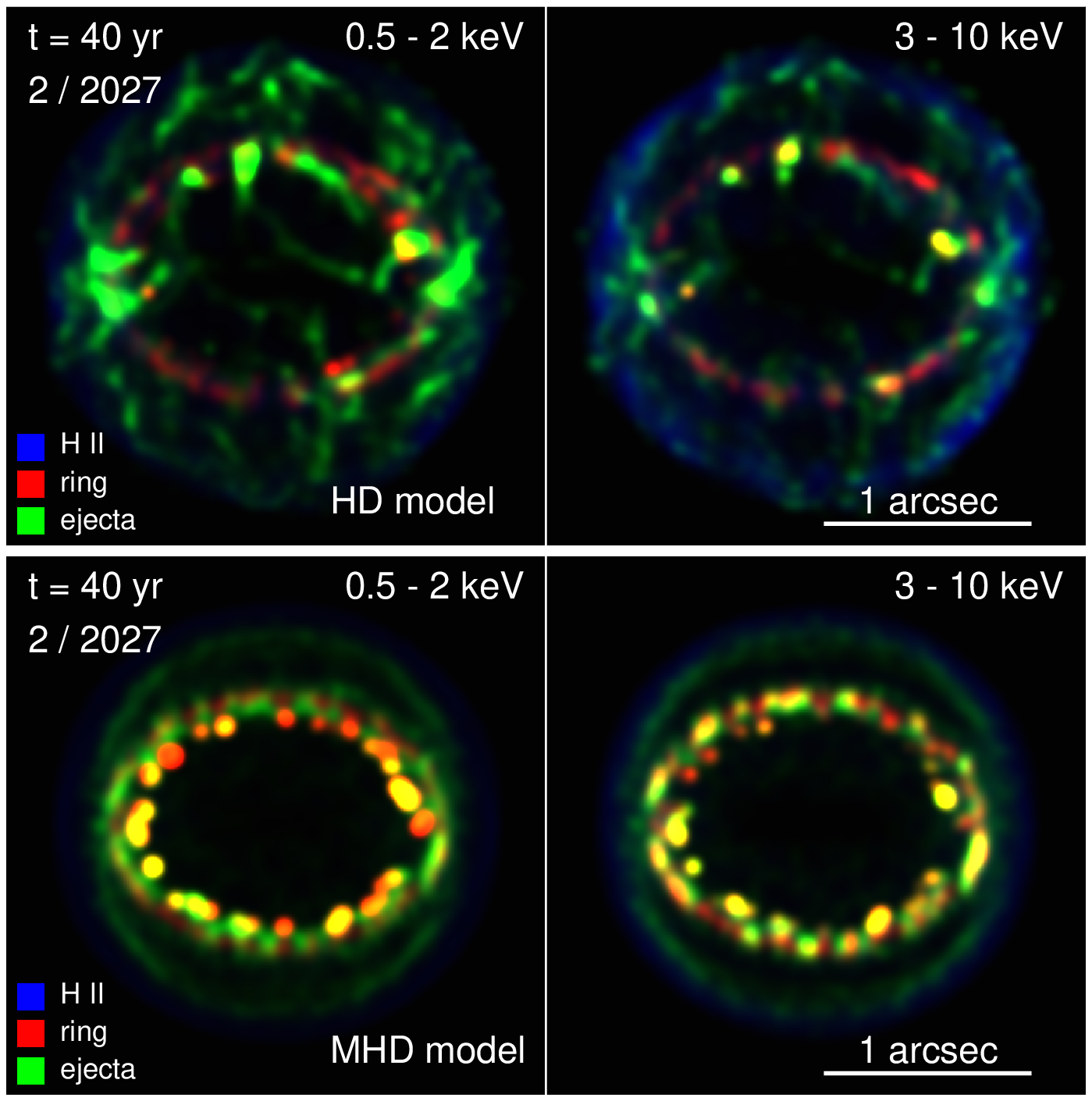, width=8.5cm}
	\caption{Three-color composite images of the X-ray emission
	in the soft ($[0.5, 2]$~keV; left panels) and hard ($[3,
	10]$~keV; right panels) bands integrated along the line of
	sight at year 40 for the HD model of \cite{2015ApJ...810..168O}
	(upper panels) and for the MHD model MOD-B1 (lower panels). Each
	image has been normalized to its maximum for visibility and
	smoothed with a Gaussian of size 0.025 arcsec. The colors
	in the composite show the contribution to emission from the
	different shocked plasma components, namely the ejecta
	(green), the ring (red), and the H II region (blue); yellow
	regions result from the combination of red and green.}
  \label{fig_emiss}
\end{center} \end{figure}

The differences in X-ray fluxes predicted by the HD and MHD models can
be explained in the light of the effects of the magnetic field in
reducing the fragmentation and destruction of the ring.
In fact, in the MHD model, the magnetic field lines gradually
envelope the dense clumps and the smooth component of the ring
during the interaction with the blast wave (see Fig.~\ref{fig_zoom}
and on-line Movie 2), thus confining efficiently the shocked plasma
of the ring and reducing the stripping of ring material by HD
instabilities. As a result, the density of the shocked ring and,
therefore, its X-ray flux are higher than in the HD model. The
ring contribution to X-ray emission is evident in Fig.~\ref{fig_emiss}
which shows three-color composite images of the X-ray emission in
the soft and hard bands integrated along the line of sight for the
HD and MHD models at year 40 (namely when the differences between
the two models are the largest). In the HD model, the ring is largely
fragmented due to the action of HD instabilities developed in the
ring surface after the passage of the blast and the emission is
dominated by shocked clumps of ejecta. In the MHD model, the ring
has survived the passage of the blast and is still evident after
40 years of evolution.

\subsection{Evolution of the radio flux}

From the model results we derived the non-thermal radio emission using
REMLIGHT, as described in Sect.~\ref{sec:radio}. The radio emissivity
in Eq.~\ref{eq6} includes the term $C'$ that was determined to fit the
observations. We found that the shape of the lightcurves assuming either
$b=0$ or $b=-1$ does not change appreciably. In the following, we discuss
in more detail the case of $b=-1$ and the same conclusions can be applied
to the case of $b=0$. We also found that, assuming either $\alpha=0.9$
or $\alpha=0.7$ (namely $\alpha$ is a constant), the synthetic lightcurves
are much flatter than observed. In the following, we discuss the case in
which $\alpha$ is given by Eq.~\ref{eq:alpha}. For our purposes, we
assumed that $C'$ in Eq.~\ref{eq6} does not depend on time (i.e. we did
not consider its dependence on the spectral index $\alpha$; see
Eq.~\ref{c_prime}).

\begin{figure*}
  \begin{center}
    \leavevmode
        \epsfig{file=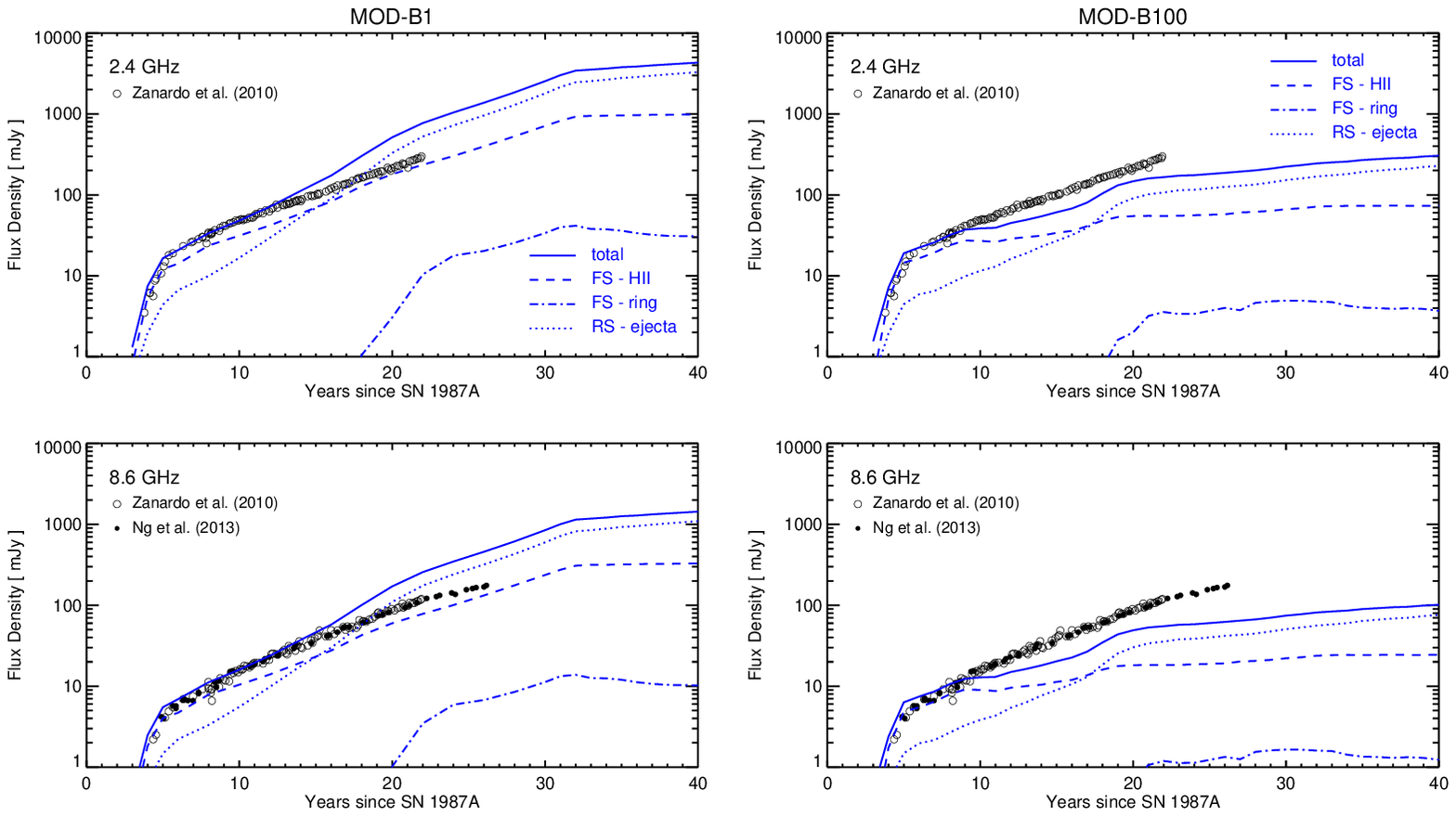, width=18.cm}
	\caption{Synthetic radio flux densities (lines) for models
	MOD-B1 (left panels) and Mod-B100 (right panels) plotted
	against the observed ones (symbols) at 2.4 GHz (upper panels;
	\citealt{2010ApJ...710.1515Z}) and 8.6 GHz (lower panels;
	\citealt{2010ApJ...710.1515Z, 2013ApJ...777..131N}). The
	figure shows the synthetic lightcurves (solid lines),
        the contribution to emission from the forward
	shock traveling through the H\,II region (dashed lines) and
	through the ring (dot-dashed lines), and the contribution from
	the reverse shock traveling through the ejecta (dotted
	lines).}
  \label{lc_radio}
\end{center} \end{figure*}

Fig.~\ref{lc_radio} shows the synthetic flux densities (solid lines)
calculated from models MOD-B1 and MOD-B100 plotted against observations
at 2.4 GHz and 8.6 GHz. During the first 13 years of evolution, the
observational data at the frequencies considered can be well reproduced
by MOD-B1: the synthetic flux densities describe the sudden rise of
radio emission occurring about three years after the SN (i.e. when the
blast wave hits the inner part of the surrounding nebula) and the almost
constant slope of the lightcurves observed in the subsequent years
when the blast wave was traveling through the H\,II region and before
its interaction with the dense ring. MOD-B100 reproduces the sudden
rise of radio emission about three years after the SN but predicts, in
the subsequent years, a slope of the lightcurves flatter than observed.
Both models fail in reproducing the observations as soon as the blast
wave hits the equatorial ring. MOD-B1 shows a significant
steepening in the radio lightcurves after the blast wave hits the ring
which is analogous to that observed in the soft X-ray lightcurve and
at odds with radio observations (left panels of Fig.~\ref{lc_radio}). A
similar discrepancy between synthetic and observed lightcurves was found
by \cite{2014ApJ...794..174P}, synthesizing the radio emission from
their 3D HD model. These authors proposed that the discrepancy might
be explained if their model overestimates the mass of the ring and/or
the ambient magnetic field within the ring. MOD-B100 shows lightcurves
flatter than in MOD-B1 and in observations, reflecting the slower decrease
of magnetic field strength with the radial distance from the center of
explosion in the equatorial plane (see Fig.~\ref{fig1}). Nevertheless,
After year 10, MOD-B100 underestimates the observed flux density at the
frequencies considered (right panels of Fig.~\ref{lc_radio}).

Thanks to the passive tracers discussed in Sect.~\ref{sec:model},
we can identify the contributions to flux density which originate
from the forward shock traveling through the H\,II region, from
that traveling through the ring, and from the reverse shock traveling
through the ejecta. Fig.~\ref{lc_radio} shows these contributions
for the two models explored. We found that most of the emission
originates from the forward shock traveling through the H\,II region
(dashed lines in Fig.~\ref{lc_radio}) and from the reverse shock
traveling through the ejecta (dotted lines). The contribution from
the shocked ring material (dot-dashed lines) is, at least, one order
of magnitude lower than that from the other two components and,
therefore, is negligible. In fact, even if the density ($n$) of the
shocked ring is much higher (by an order of magnitude, on average)
than that of the shocked plasma from the H\,II region, the volume
($V$) occupied by this plasma component is much lower (by a factor
of $\approx 30$) than that of the shocked H\,II region. This together
with the fact that the radio flux $F\rs{radio}\propto n V$, whereas
the X-ray flux $F\rs{X-ray}\propto n^2 V$, explain why the radio
lightcurves are not dominated by emission from the shocked
ring, at variance with the soft X-ray lightcurve which indeed is
dominated by emission from the shocked ring material (cf.
Fig.~\ref{fig_lc}). Thus our results rule out the possibility that
the models may overestimate the mass of the ring and/or the ambient
magnetic field within the ring as suggested by \cite{2014ApJ...794..174P}.

By inspecting Fig.~\ref{lc_radio}, we note that a magnetic field
configuration intermediate between those considered here might fit the
observed lightcurves. In this case the radio emission may be dominated
by the H\,II region during the first $\approx 15$ years of evolution
and by the shocked ejecta at later times. On the other hand,
Fig.~\ref{lc_radio} also shows that, in model MOD-B1, the steepening
in the radio lightcurves around year 13 is caused by the contribution
from the shocked ejecta (namely from the reverse shock) which becomes
dominant about 15 years after the SN event, namely after the blast
wave hits the equatorial ring. If this contribution were suppressed
(at least by $\approx 2$ orders of magnitude) the observational
data at both frequencies might be well reproduced by the model (see
Fig.~\ref{lc_radio_sup}). In this case, the emission would be entirely
dominated by the forward shock traveling through the H\,II region (the
contribution from the shocked ring is orders of magnitude lower). The
model would describe the sudden rise of radio emission occurring about
three years after the SN and the almost constant slope of the lightcurves
observed in the subsequent 27 years of evolution.

We note that \cite{2012APh....35..300T} have investigated the
particle acceleration at forward and reverse shocks of young type
Ia SNRs and found that reverse shock contribution to the cosmic-ray
particle population may be significant. However, a dominant radio
emission from the shocked H\,II region (and a negligible contribution
from the shocked ring and from the reverse shock) is an interesting
possibility because it might explain why the radio remnant (which,
according to our model, shows the expansion of the blast wave through
the H\,II region) expands almost linearly up to year 22
(\citealt{2008ApJ...684..481N, 2013ApJ...777..131N}), in contrast
with the X-ray remnant which shows a clear deceleration in the
expansion around year 16 (\citealt{2013ApJ...764...11H,
2016ApJ...829...40F}), namely when the X-ray emission becomes
dominated by the emission from the shocked ring
(\citealt{2015ApJ...810..168O}; see Fig.~\ref{fig_lc}). Also, this
might explain why \cite{2008ApJ...684..481N} are able to fit the
radio data more accurately by a torus model which can capture the
latitude extent of the emission (due to the thickness of the H\,II
region) rather than by a ring model (which is thinner), and why the
radio remnant, dominated by emission from the shocked H\,II region,
appears to be larger than the X-ray remnant which is dominated by
emission from the slower shock traveling through the equatorial
ring (\citealt{2018arXiv180902364C}).
The above scenario is also supported by the findings of
\cite{2017A&A...605A.110P} who found that the thickness of the H\,II
region in the 3D HD model of \cite{2015ApJ...810..168O} fits well the
evolution of the latitude extent of the radio emission (see their Fig.~6)
derived by \cite{2008ApJ...684..481N, 2013ApJ...777..131N}.

\subsection{A mechanism to suppress radio emission from the reverse
shock?}

In the light of the above discussion, a mechanism to suppress
radio emission from the reverse shock may be invoked to reconcile
models and observations. A hint to a possible mechanism comes
from observations of the radio emission from SN\,1987A: a number
of authors inferred SSA to occur in the early expanding remnant of
SN\,1987A (e.g. \citealt{1987Natur.329..421S, 1992A&A...254..167K,
1998ApJ...499..810C}). Thus we investigated if SSA or FFA might still
play a role during the later interaction of the remnant with the nebula
by using the 1D CR-hydro-NEI code (\citealt{2012ApJ...750..156L},
see Sect.~\ref{sec:radio}). As initial conditions, we considered the
radial profiles of density, pressure, velocity, and magnetic field
strength derived from model MOD-B1 in the plane of the ring about one
year after the SN event (namely before the interaction of the blast
wave with the nebula). We evolved the system for 30 years, deriving the
radio lightcurve in the 2.4 GHz. We found that SSA does not play any
significant role in the model in the period between years 3 and 30. This
is mainly due to the fact that the magnetic field is never remarkably
large both for the shocked ejecta and CSM. FFA does
cut off the lower frequencies for both ejecta and CSM emission at a
comparable level, but it does not affect the spectrum around 2.4 GHz
(and at larger frequencies). This is in agreement also with previous
studies on the role of SSA and FFA in the late radio emission from SN
explosions (e.g. \citealt{2001A&A...374..997P}): at frequencies higher
than 1 GHz these effects are negligible after $\sim 1400\ {\rm days}\sim
4\ {\rm yrs}$ since the SN.

\begin{figure}
  \begin{center}
    \leavevmode
        \epsfig{file=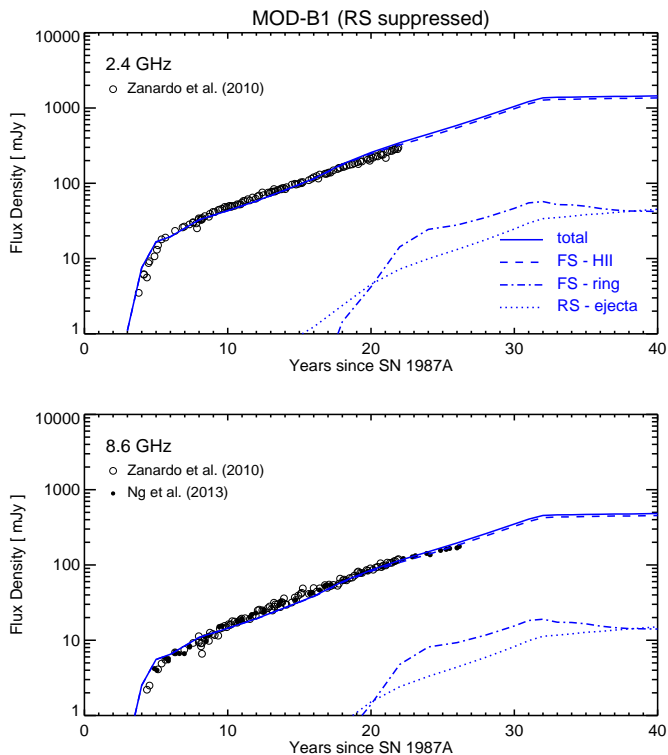, width=9.cm}
        \caption{As in Fig.~\ref{lc_radio} for model MOD-B1 and assuming
        the emission from the reverse shock reduced by two orders
        of magnitude.}
  \label{lc_radio_sup}
\end{center}
\end{figure}

The 1D CR-hydro-NEI simulations show that a very strong boost of radio
emissivity from the ejecta is resulted from their interaction with the
nebula (to a larger extent than the forward shock emission). Following a
strong compression of the shocked ejecta due to the rapid deceleration
of the forward shock and to the feedback of accelerated particles
(which is included in CR-hydro-NEI calculations), the reverse shock
is highly strengthened and pushed into the high density envelope of
the ejecta, resulting into a rapid increase of accelerated electrons
in the ejecta. This explains why, if the diffusive shock acceleration
(DSA) efficiency at the reverse shock is not suppressed, the radio flux
from the ejecta can easily overshoot the observed flux as also found
with the synthesis of radio emission from the 3D MHD simulations (see
Fig.~\ref{lc_radio}). On the contrary, if the DSA at the reverse shock
is highly suppressed, the model can reproduce the observed lightcurves,
thus strongly favoring a shocked CSM (forward shock) origin of the radio
emission, in agreement with the findings of Fig.~\ref{lc_radio_sup}. In
such a case, the acceleration efficiency at the reverse shock might be
strongly constrained.

\begin{figure*}
  \begin{center}
    \leavevmode
        \epsfig{file=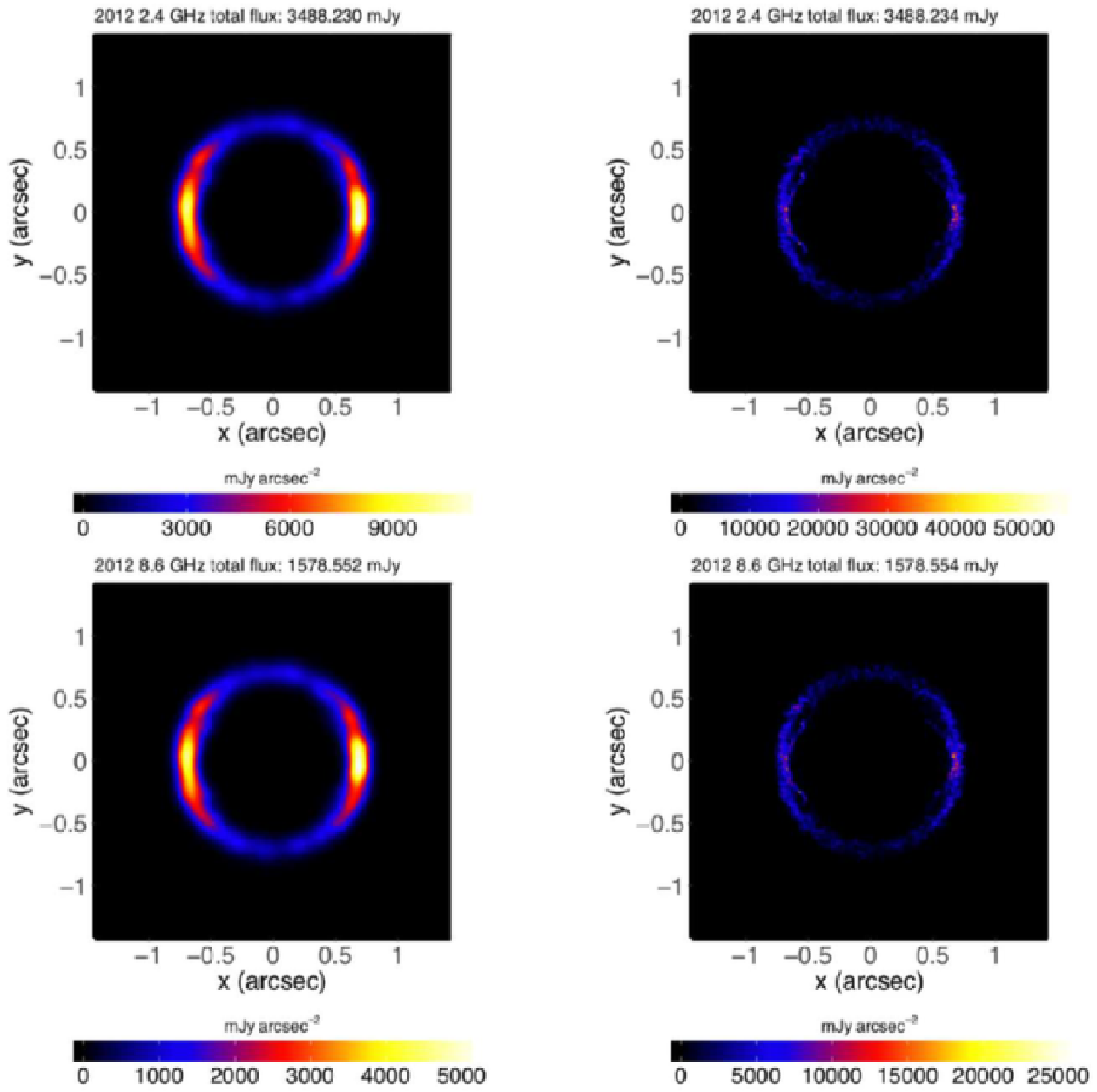, width=18.cm}
        \caption{Synthetic maps at 2.4 GHz (upper panels) and 8.6 GHz
        (lower panels) for the 2012 epoch (24 years after the SN) from
        the model MOD-B1. The left column shows the images convolved
        by a beam with FWHM of 0.5 arc seconds, while the right column
        shows the unconvolved images. The total flux in convolved images is
        the same of that in unconvolved images.}
  \label{img_radio}
\end{center}
\end{figure*}

The CR-hydro-NEI code assumes spherical symmetry in the expansion of
the remnant and it cannot describe the inhomogeneous structure of the
nebula around SN\,1987A. In particular it does not describe the complex
configuration of the magnetic field when the blast wave runs over the
dense ring (see Sect.~\ref{sec:field}). Thus, to investigate further
the role played by SSA in the radio evolution of SN\,1987A, we used the
SPEV code (\citealt{2009ApJ...696.1142M}; see Sect.~\ref{sec:radio})
to synthesize, from run MOD-B1, the radio emission at 2.4 GHz and 8.6
GHz. Fig.~\ref{img_radio} shows an example of radio maps obtained with
SPEV for year 24 (i.e. the 2012 epoch), assuming isotropic injection. We
found that some absorption of emission from the reverse shock can be
present in correspondence of the equatorial ring, especially at 2.4
GHz. This is mainly due to the dense material of clumps with
number density $\approx 10^4$~cm$^{-3}$ and temperature around
$10^4$~K. Nevertheless this absorption reduces the emission from the
reverse shock by only a few percents and it is not able to account for
a significant suppression of the radio flux.

In the synthesis of radio emission, we have assumed that the
parameter $C'$ is the same for forward and reverse shocks ($C'\rs{FS}
= C'\rs{RS}$). This parameter describes the contributions to the radio
emission from electron populations accelerated by the shock. However,
as shown in Fig.~\ref{lc_radio_sup}, the radio lightcurves 
can be reproduced by our model if the DSA at the reverse shock is
highly suppressed. The parameter $C'$ is linked to the number of
particles available for the acceleration and to the magnetization of
the pre-shock medium. In the case of SN\,1987A, many authors suggest
that the unshocked ejecta should be mainly in neutral or in singly
ionized states (e.g. \citealt{2011A&A...530A..45J}). In this case, very
few particles should participate in the process of acceleration at the
reverse shock. However, once the ejecta is crossed by the reverse shock,
the material is heated up to temperatures of several millions degree and
ionized, possibly introducing fresh particles available for acceleration.  It is
more likely, therefore, that $C'\rs{RS} << C'\rs{FS}$ due to a different
magnetization of plasma. In fact, first principle particle-in-cell
(PIC) simulations of perpendicular strongly magnetized shocks have
shown that accelerated electrons can be heavily suppressed as
a consequence of the lack of sufficient self-generated turbulence
(e.g. \citealt{2013ApJ...771...54S}).

According to \cite{2013ApJ...771...54S}, the maximum Lorentz factor at
saturation for the electrons, $\gamma_{sat,e}$, in a relativistic
magnetized shock in an electron-proton plasma is

\begin{equation}
\gamma_{sat,e} \approx 2 \gamma_0 \frac{m_i}{m_e} \sigma_{m}^{-1/4},
\end{equation}

\noindent
where $\gamma_0$ is the Lorentz factor, $m_i$ and $m_e$ are the
masses of ions and electrons, respectively, and $\sigma_m\approx 50
B^2 n^{-1}$ is the magnetization calculated as the ratio of magnetic
to kinetic energy density. \cite{2013ApJ...771...54S} have shown that,
with increasing magnetization, the acceleration efficiency decreases: if
the magnetization in the pre-shock is larger than $\approx 10^{-4}$, the
post-shock spectrum does not show any evidence for non-thermal particles
and the radiation in the bands of interest is strongly suppressed. In the
case of SN\,1987A (which is non-relativistic), we expect that the critical
value of $\sigma_{m}$ at which the radio emission is suppressed might be
different. Nevertheless, if the unshocked ejecta is more magnetized
than the CSM, and if it is above some critical value, this could be an
argument in favor of suppression of the reverse shock emission. However,
in model MOD-B1, the reverse shock magnetization is lower than that of
the forward shock. Also $\sigma_m < 5\times 10^{-6}$ in the ejecta, too
low to make a difference in $\gamma_{sat,e}$ in enough emitting volumes;
in fact we found that the difference in flux is at the level of 0.1\%.

On the other hand, $\sigma_m$ increases as $B^2$. For instance, by
increasing $B$ in the unshocked ejecta of our model by a factor of 100,
$\sigma_m$ should be a factor of $10^4$ larger. Thus more magnetized
models (especially in the ejecta) would yield a limit for the maximum
Lorentz factor that would forbid the emission at radio frequencies in
the reverse shock. We suggest, therefore, that the model considered
here underestimates the magnetic field strength in the ejecta so that
the radio emission is not suppressed at the reverse shock. 
\cite{2018ApJ...861L...9Z} found a magnetic field strength of a few mG
in the region between the forward and reverse shocks in SN\,1987A. If the
magnetic field strength in the ejecta is of the order of that inferred
from the observations, $\sigma_m$ might large enough to provide a natural
explanation for the suppression of radio emission from the reverse shock.

We note that our model assumes a magnetic field strength and
configuration in the initial remnant interior (at $t\approx 24$~hours
since the SN) which are derived from the Parker spiral model (which
is appropriate for the CSM). We expect, however, that a more realistic
field (especially in the immediate surroundings of the remnant
compact object) would be much more complex than that adopted here
and it should reflect the field of the stellar interior before the
collapse of the stellar progenitor. Unfortunately we do not have
any hints on this magnetic field. Also, it is likely that
efficient field amplification by MRI occurred after core bounce
(e.g. \citealt{2018JPhG...45h4001O}). We conclude that, most likely,
we are largely underestimating the strength of this magnetic field.

\section{Summary and conclusions}
\label{sec:conc}

We modeled the evolution of SN\,1987A with the aim to investigate
the role played by a pre-SN ambient magnetic field in the dynamics
of the expanding remnant and to ascertain the origin of its radio
emission. To this end, we developed a 3D MHD model describing the
evolution of SN\,1987A  from the immediate aftermath of the SN
explosion to the development of the full-fledged remnant, covering
40 years of evolution since few hours after the SN event. The model
considers a plausible configuration of the pre-SN ambient magnetic
field and includes the deviations from equilibrium of ionization,
and the deviation from temperature-equilibration between electrons and
ions. The initial condition is the output of a SN model able to
reproduce the bolometric lightcurve of SN\,1987A during the first
250 days of evolution. The nebula around SN\,1987A is modeled as
in \cite{2015ApJ...810..168O}, adopting the parameters of the CSM
derived by these authors to fit the X-ray observations (morphology,
lightcurves and spectra) of the remnant during the first 30 years
of evolution. Our findings can be summarized as follows.

\begin{enumerate}
\item The presence of an ambient magnetic field does not change
significantly the overall evolution and the morphology of the
remnant. The 3D MHD model reproduces the X-ray morphology and
lightcurves of SN\,1987A during the first 30 years of evolution,
by adopting the same initial condition (the SN model) and boundary
conditions (the geometry and density distribution of the nebula
around SN\,1987A) of the best-fit HD model of SN\,1987A presented
in \cite{2015ApJ...810..168O}.

\item The magnetic field plays a significant role in reducing the
erosion and fragmentation of the dense equatorial ring after the
passage of the SN blast wave. In particular, the field maintains a
more laminar flow around the ring, limiting the growth of HD
instabilities that would develop at the ring boundary. As a
consequence, at variance with the results from the HD model, the
ring survives the passage of the blast until the end of the simulation
(40 years). The ring contribution to thermal X-ray emission is
slightly higher than that in the HD model and is the dominant
component until the end of the simulation (40~years after the SN
event).

\item The two models explored (MOD-B1 and MOD-B100) predict different
slopes of the radio lightcurves due to different radial variations of
the magnetic field strength considered in the two cases. MOD-B1 is able
to fit the radio lightcurves of SN\,1987A during the first 13 years of
evolution, namely when the blast wave travels through the H\,II region. In
that period most of the radio emission originates from the forward shock
traveling through the H\,II region. At later times, the model predicts a
significant contribution to the radio flux originating from the reverse
shock. This contribution becomes dominant after the blast wave hits
the equatorial ring (about 15 years after the SN event). However, this
contribution leads to a steepening in the radio lightcurves at odds with
observations. MOD-B100 predicts a slope of the lightcurves flatter
than that observed already at early times ($\approx 5$ years after the SN)
and it is not able to fit the observed lightcurves. In both models, the
radio flux arising from the shocked ring material is negligible because
the volume occupied by this plasma component is much smaller than that
of the shocked H\,II region.

\item MOD-B1 may reproduce the radio lightcurves over the whole
observed period if the flux originating from the reverse shock
was heavily suppressed (by $\approx 2$ orders of magnitude; see
Fig.~\ref{lc_radio_sup}). We explored if SSA and/or FFA could be possible
candidates for the absorption mechanism of the radio emission and we
found that they do not play any significant role in the model after 3
years since the SN. On the other hand, the magnetic field strength and
configuration adopted in our model are arbitrary (the only constraint
is the field strength at the inner edge of the nebula), especially in
the remnant interior where the field should reflect that of the stellar
interior before the collapse of the progenitor star. We suggest that
a larger magnetic field in the unshocked ejecta would yield a limit
for the maximum Lorentz factor that would forbid the emission at radio
frequencies in the reverse shock.

\end{enumerate}

\begin{acknowledgements}

We acknowledge that the results of this research have been achieved
using the PRACE Research Infrastructure resource Marconi based in
Italy at CINECA (PRACE Award N.2016153460). The PLUTO code, used
in this work, is developed at the Turin Astronomical Observatory
in collaboration with the Department of General Physics of Turin
University and the SCAI Department of CINECA. SO thanks Andrea
Mignone for his help and support with the PLUTO code. SO, MM, GP, FB
acknowledge financial contribution from the agreement ASI-INAF
n.2017-14-H.O, and partial financial support by the PRIN INAF 2016
grant ``Probing particle acceleration and $\gamma$-ray propagation
with CTA and its precursors''. SHL acknowledges support by the Kyoto
University Foundation. SN wishes to acknowledge the support of the
Program of Interdisciplinary Theoretical \& Mathematical Science
(iTHEMS) at RIKEN. MAA and PM acknowledge the support of the Spanish
Ministry of Economy and Competitiveness through grant AYA2015-66899-C2-1-P
and the partial support through the PHAROS COST Action (CA16214)
and the GWverse COST Action (CA16104).

\end{acknowledgements}

\bibliographystyle{aa}
\bibliography{biblio}

\begin{thebibliography}{89}
\expandafter\ifx\csname natexlab\endcsname\relax\def\natexlab#1{#1}\fi

\bibitem[{{Cendes} {et~al.}(2018){Cendes}, {Gaensler}, {Ng}, {Zanardo},
  {Staveley-Smith}, \& {Tzioumis}}]{2018arXiv180902364C}
{Cendes}, Y., {Gaensler}, B., {Ng}, C.-Y., {et~al.} 2018, ArXiv e-prints

\bibitem[{{Chevalier}(1998)}]{1998ApJ...499..810C}
{Chevalier}, R.~A. 1998, \apj, 499, 810

\bibitem[{{Chevalier} \& {Dwarkadas}(1995)}]{1995ApJ...452L..45C}
{Chevalier}, R.~A. \& {Dwarkadas}, V.~V. 1995, Astrophys. J., 452, L45

\bibitem[{{Chita} {et~al.}(2008){Chita}, {Langer}, {van Marle},
  {Garc{\'{\i}}a-Segura}, \& {Heger}}]{2008A&A...488L..37C}
{Chita}, S.~M., {Langer}, N., {van Marle}, A.~J., {Garc{\'{\i}}a-Segura}, G.,
  \& {Heger}, A. 2008, \aap, 488, L37

\bibitem[{{Crotts} \& {Heathcote}(2000)}]{2000ApJ...528..426C}
{Crotts}, A.~P.~S. \& {Heathcote}, S.~R. 2000, \apj, 528, 426

\bibitem[{{Crotts} {et~al.}(1989){Crotts}, {Kunkel}, \&
  {McCarthy}}]{1989ApJ...347L..61C}
{Crotts}, A.~P.~S., {Kunkel}, W.~E., \& {McCarthy}, P.~J. 1989, \apjl, 347, L61

\bibitem[{{Dedner} {et~al.}(2002){Dedner}, {Kemm}, {Kr{\"o}ner}, {Munz},
  {Schnitzer}, \& {Wesenberg}}]{2002JCoPh.175..645D}
{Dedner}, A., {Kemm}, F., {Kr{\"o}ner}, D., {et~al.} 2002, Journal of
  Computational Physics, 175, 645

\bibitem[{{Dickel} \& {Milne}(1976)}]{1976AuJPh..29..435D}
{Dickel}, J.~R. \& {Milne}, D.~K. 1976, Australian Journal of Physics, 29, 435

\bibitem[{{Donati} \& {Landstreet}(2009)}]{2009ARA&A..47..333D}
{Donati}, J.-F. \& {Landstreet}, J.~D. 2009, \araa, 47, 333

\bibitem[{{Dubner} \& {Giacani}(2015)}]{2015A&ARv..23....3D}
{Dubner}, G. \& {Giacani}, E. 2015, \aapr, 23, 3

\bibitem[{{Fragile} {et~al.}(2005){Fragile}, {Anninos}, {Gustafson}, \&
  {Murray}}]{2005ApJ...619..327F}
{Fragile}, P.~C., {Anninos}, P., {Gustafson}, K., \& {Murray}, S.~D. 2005,
  \apj, 619, 327

\bibitem[{{Frank} {et~al.}(2016){Frank}, {Zhekov}, {Park}, {McCray}, {Dwek}, \&
  {Burrows}}]{2016ApJ...829...40F}
{Frank}, K.~A., {Zhekov}, S.~A., {Park}, S., {et~al.} 2016, \apj, 829, 40

\bibitem[{{Gaensler} {et~al.}(1997){Gaensler}, {Manchester}, {Staveley-Smith},
  {Tzioumis}, {Reynolds}, \& {Kesteven}}]{1997ApJ...479..845G}
{Gaensler}, B.~M., {Manchester}, R.~N., {Staveley-Smith}, L., {et~al.} 1997,
  \apj, 479, 845

\bibitem[{{Gawryszczak} {et~al.}(2010){Gawryszczak}, {Guzman}, {Plewa}, \&
  {Kifonidis}}]{2010A&A...521A..38G}
{Gawryszczak}, A., {Guzman}, J., {Plewa}, T., \& {Kifonidis}, K. 2010, \aap,
  521, A38

\bibitem[{{Ghavamian} {et~al.}(2007){Ghavamian}, {Laming}, \&
  {Rakowski}}]{2007ApJ...654L..69G}
{Ghavamian}, P., {Laming}, J.~M., \& {Rakowski}, C.~E. 2007, \apjl, 654, L69

\bibitem[{{Ginzburg} \& {Syrovatskii}(1965)}]{1965ARA&A...3..297G}
{Ginzburg}, V.~L. \& {Syrovatskii}, S.~I. 1965, \araa, 3, 297

\bibitem[{{Haberl} {et~al.}(2006){Haberl}, {Geppert}, {Aschenbach}, \&
  {Hasinger}}]{2006A&A...460..811H}
{Haberl}, F., {Geppert}, U., {Aschenbach}, B., \& {Hasinger}, G. 2006, \aap,
  460, 811

\bibitem[{{Hasinger} {et~al.}(1996){Hasinger}, {Aschenbach}, \&
  {Truemper}}]{1996A&A...312L...9H}
{Hasinger}, G., {Aschenbach}, B., \& {Truemper}, J. 1996, \aap, 312, L9

\bibitem[{{Heger} {et~al.}(2005){Heger}, {Woosley}, \&
  {Spruit}}]{2005ApJ...626..350H}
{Heger}, A., {Woosley}, S.~E., \& {Spruit}, H.~C. 2005, \apj, 626, 350

\bibitem[{{Helder} {et~al.}(2013){Helder}, {Broos}, {Dewey}, {Dwek}, {McCray},
  {Park}, {Racusin}, {Zhekov}, \& {Burrows}}]{2013ApJ...764...11H}
{Helder}, E.~A., {Broos}, P.~S., {Dewey}, D., {et~al.} 2013, Astrophys. J.,
  764, 11

\bibitem[{{Hillebrandt} {et~al.}(1987){Hillebrandt}, {Hoeflich}, {Weiss}, \&
  {Truran}}]{1987Natur.327..597H}
{Hillebrandt}, W., {Hoeflich}, P., {Weiss}, A., \& {Truran}, J.~W. 1987, \nat,
  327, 597

\bibitem[{{Hole} {et~al.}(2010){Hole}, {Kasen}, \&
  {Nordsieck}}]{2010ApJ...720.1500H}
{Hole}, K.~T., {Kasen}, D., \& {Nordsieck}, K.~H. 2010, \apj, 720, 1500

\bibitem[{{Jerkstrand} {et~al.}(2011){Jerkstrand}, {Fransson}, \&
  {Kozma}}]{2011A&A...530A..45J}
{Jerkstrand}, A., {Fransson}, C., \& {Kozma}, C. 2011, \aap, 530, A45

\bibitem[{{Jones} {et~al.}(1996){Jones}, {Ryu}, \&
  {Tregillis}}]{1996ApJ...473..365J}
{Jones}, T.~W., {Ryu}, D., \& {Tregillis}, I.~L. 1996, \apj, 473, 365

\bibitem[{{Kashyap} \& {Drake}(2000)}]{Kashyap2000BASI}
{Kashyap}, V. \& {Drake}, J.~J. 2000, Bulletin of the Astronomical Society of
  India, 28, 475

\bibitem[{{Kifonidis} {et~al.}(2006){Kifonidis}, {Plewa}, {Scheck}, {Janka}, \&
  {M{\"u}ller}}]{2006A&A...453..661K}
{Kifonidis}, K., {Plewa}, T., {Scheck}, L., {Janka}, H.-T., \& {M{\"u}ller}, E.
  2006, \aap, 453, 661

\bibitem[{{Kirk} \& {Wassmann}(1992)}]{1992A&A...254..167K}
{Kirk}, J.~G. \& {Wassmann}, M. 1992, \aap, 254, 167

\bibitem[{{Lee} {et~al.}(2012){Lee}, {Ellison}, \&
  {Nagataki}}]{2012ApJ...750..156L}
{Lee}, S.-H., {Ellison}, D.~C., \& {Nagataki}, S. 2012, \apj, 750, 156

\bibitem[{{Mac Low} {et~al.}(1994){Mac Low}, {McKee}, {Klein}, {Stone}, \&
  {Norman}}]{1994ApJ...433..757M}
{Mac Low}, M., {McKee}, C.~F., {Klein}, R.~I., {Stone}, J.~M., \& {Norman},
  M.~L. 1994, \apj, 433, 757

\bibitem[{{Maggi} {et~al.}(2012){Maggi}, {Haberl}, {Sturm}, \&
  {Dewey}}]{2012A&A...548L...3M}
{Maggi}, P., {Haberl}, F., {Sturm}, R., \& {Dewey}, D. 2012, Astron. \&
  Astrophys., 548, L3

\bibitem[{{Masada} {et~al.}(2018){Masada}, {Kotake}, {Takiwaki}, \&
  {Yamamoto}}]{2018PhRvD..98h3018M}
{Masada}, Y., {Kotake}, K., {Takiwaki}, T., \& {Yamamoto}, N. 2018, \prd, 98,
  083018

\bibitem[{{Masada} {et~al.}(2012){Masada}, {Takiwaki}, {Kotake}, \&
  {Sano}}]{2012ApJ...759..110M}
{Masada}, Y., {Takiwaki}, T., {Kotake}, K., \& {Sano}, T. 2012, \apj, 759, 110

\bibitem[{{McCray}(2007)}]{2007AIPC..937....3M}
{McCray}, R. 2007, in American Institute of Physics Conference Series, Vol.
  937, Supernova 1987A: 20 Years After: Supernovae and Gamma-Ray Bursters, ed.
  S.~{Immler}, K.~{Weiler}, \& R.~{McCray}, 3--14

\bibitem[{{McCray} \& {Fransson}(2016)}]{2016ARA&A..54...19M}
{McCray}, R. \& {Fransson}, C. 2016, \araa, 54, 19

\bibitem[{{Mignone} {et~al.}(2007){Mignone}, {Bodo}, {Massaglia}, {Matsakos},
  {Tesileanu}, {Zanni}, \& {Ferrari}}]{2007ApJS..170..228M}
{Mignone}, A., {Bodo}, G., {Massaglia}, S., {et~al.} 2007, \apjs, 170, 228

\bibitem[{{Mignone} {et~al.}(2010){Mignone}, {Tzeferacos}, \&
  {Bodo}}]{2010JCoPh.229.5896M}
{Mignone}, A., {Tzeferacos}, P., \& {Bodo}, G. 2010, Journal of Computational
  Physics, 229, 5896

\bibitem[{{Mignone} {et~al.}(2012){Mignone}, {Zanni}, {Tzeferacos}, {van
  Straalen}, {Colella}, \& {Bodo}}]{2012ApJS..198....7M}
{Mignone}, A., {Zanni}, C., {Tzeferacos}, P., {et~al.} 2012, \apjs, 198, 7

\bibitem[{{Mimica} {et~al.}(2009){Mimica}, {Aloy}, {Agudo}, {Mart{\'{\i}}},
  {G{\'o}mez}, \& {Miralles}}]{2009ApJ...696.1142M}
{Mimica}, P., {Aloy}, M.-A., {Agudo}, I., {et~al.} 2009, \apj, 696, 1142

\bibitem[{{Miyoshi} \& {Kusano}(2005)}]{2005JCoPh.208..315M}
{Miyoshi}, T. \& {Kusano}, K. 2005, Journal of Computational Physics, 208, 315

\bibitem[{{Morris} \& {Podsiadlowski}(2007)}]{2007Sci...315.1103M}
{Morris}, T. \& {Podsiadlowski}, P. 2007, Science, 315, 1103

\bibitem[{{Nagataki}(2000)}]{2000ApJS..127..141N}
{Nagataki}, S. 2000, \apjs, 127, 141

\bibitem[{{Nagataki} {et~al.}(1997){Nagataki}, {Hashimoto}, {Sato}, \&
  {Yamada}}]{1997ApJ...486.1026N}
{Nagataki}, S., {Hashimoto}, M.-a., {Sato}, K., \& {Yamada}, S. 1997, \apj,
  486, 1026

\bibitem[{{Ng} {et~al.}(2008){Ng}, {Gaensler}, {Staveley-Smith}, {Manchester},
  {Kesteven}, {Ball}, \& {Tzioumis}}]{2008ApJ...684..481N}
{Ng}, C.-Y., {Gaensler}, B.~M., {Staveley-Smith}, L., {et~al.} 2008, \apj, 684,
  481

\bibitem[{{Ng} {et~al.}(2013){Ng}, {Zanardo}, {Potter}, {Staveley-Smith},
  {Gaensler}, {Manchester}, \& {Tzioumis}}]{2013ApJ...777..131N}
{Ng}, C.-Y., {Zanardo}, G., {Potter}, T.~M., {et~al.} 2013, \apj, 777, 131

\bibitem[{{Obergaulinger} \& {Aloy}(2017)}]{2017MNRAS.469L..43O}
{Obergaulinger}, M. \& {Aloy}, M.~{\'A}. 2017, \mnras, 469, L43

\bibitem[{{Obergaulinger} {et~al.}(2015){Obergaulinger}, {Chimeno}, {Mimica},
  {Aloy}, \& {Iyudin}}]{2015HEDP...17...92O}
{Obergaulinger}, M., {Chimeno}, J.~M., {Mimica}, P., {Aloy}, M.~A., \&
  {Iyudin}, A. 2015, High Energy Density Physics, 17, 92

\bibitem[{{Obergaulinger} {et~al.}(2014){Obergaulinger}, {Janka}, \&
  {Aloy}}]{2014MNRAS.445.3169O}
{Obergaulinger}, M., {Janka}, H.-T., \& {Aloy}, M.~A. 2014, \mnras, 445, 3169

\bibitem[{{Obergaulinger} {et~al.}(2018){Obergaulinger}, {Just}, \&
  {Aloy}}]{2018JPhG...45h4001O}
{Obergaulinger}, M., {Just}, O., \& {Aloy}, M.~A. 2018, Journal of Physics G
  Nuclear Physics, 45, 084001

\bibitem[{{Ono} {et~al.}(2013){Ono}, {Nagataki}, {Ito}, {Lee}, {Mao},
  {Hashimoto}, \& {Tolstov}}]{2013ApJ...773..161O}
{Ono}, M., {Nagataki}, S., {Ito}, H., {et~al.} 2013, \apj, 773, 161

\bibitem[{{Orlando} {et~al.}(2006){Orlando}, {Peres}, {Reale}, {Plewa}, \&
  {Rosner}}]{orlando2}
{Orlando}, S.~{Bocchino}, F., {Peres}, G., {Reale}, F., {Plewa}, T., \&
  {Rosner}, R. 2006, \aap, 457, 545

\bibitem[{{Orlando} {et~al.}(2012){Orlando}, {Bocchino}, {Miceli}, {Petruk}, \&
  {Pumo}}]{2012ApJ...749..156O}
{Orlando}, S., {Bocchino}, F., {Miceli}, M., {Petruk}, O., \& {Pumo}, M.~L.
  2012, \apj, 749, 156

\bibitem[{{Orlando} {et~al.}(2008){Orlando}, {Bocchino}, {Reale}, {Peres}, \&
  {Pagano}}]{2008ApJ...678..274O}
{Orlando}, S., {Bocchino}, F., {Reale}, F., {Peres}, G., \& {Pagano}, P. 2008,
  \apj, 678, 274

\bibitem[{{Orlando} {et~al.}(2007){Orlando}, {Bocchino}, {Reale}, {Peres}, \&
  {Petruk}}]{2007A&A...470..927O}
{Orlando}, S., {Bocchino}, F., {Reale}, F., {Peres}, G., \& {Petruk}, O. 2007,
  \aap, 470, 927

\bibitem[{{Orlando} {et~al.}(2009){Orlando}, {Drake}, \&
  {Laming}}]{2009A&A...493.1049O}
{Orlando}, S., {Drake}, J.~J., \& {Laming}, J.~M. 2009, \aap, 493, 1049

\bibitem[{{Orlando} {et~al.}(2015){Orlando}, {Miceli}, {Pumo}, \&
  {Bocchino}}]{2015ApJ...810..168O}
{Orlando}, S., {Miceli}, M., {Pumo}, M.~L., \& {Bocchino}, F. 2015, \apj, 810,
  168

\bibitem[{{Orlando} {et~al.}(2016){Orlando}, {Miceli}, {Pumo}, \&
  {Bocchino}}]{2016ApJ...822...22O}
{Orlando}, S., {Miceli}, M., {Pumo}, M.~L., \& {Bocchino}, F. 2016, \apj, 822,
  22

\bibitem[{{Orlando} {et~al.}(2011){Orlando}, {Petruk}, {Bocchino}, \&
  {Miceli}}]{2011A&A...526A.129O}
{Orlando}, S., {Petruk}, O., {Bocchino}, F., \& {Miceli}, M. 2011, \aap, 526,
  A129

\bibitem[{{Panagia}(1999)}]{1999IAUS..190..549P}
{Panagia}, N. 1999, in IAU Symposium, Vol. 190, New Views of the Magellanic
  Clouds, ed. Y.-H. {Chu}, N.~{Suntzeff}, J.~{Hesser}, \& D.~{Bohlender}, 549

\bibitem[{{Park} {et~al.}(2006){Park}, {Zhekov}, {Burrows}, {Garmire},
  {Racusin}, \& {McCray}}]{2006ApJ...646.1001P}
{Park}, S., {Zhekov}, S.~A., {Burrows}, D.~N., {et~al.} 2006, \apj, 646, 1001

\bibitem[{{Park} {et~al.}(2005){Park}, {Zhekov}, {Burrows}, \&
  {McCray}}]{2005ApJ...634L..73P}
{Park}, S., {Zhekov}, S.~A., {Burrows}, D.~N., \& {McCray}, R. 2005, \apjl,
  634, L73

\bibitem[{{Parker}(1958)}]{1958ApJ...128..664P}
{Parker}, E.~N. 1958, \apj, 128, 664

\bibitem[{{P{\'e}rez-Torres} {et~al.}(2001){P{\'e}rez-Torres}, {Alberdi}, \&
  {Marcaide}}]{2001A&A...374..997P}
{P{\'e}rez-Torres}, M.~A., {Alberdi}, A., \& {Marcaide}, J.~M. 2001, \aap, 374,
  997

\bibitem[{{Petermann} {et~al.}(2015){Petermann}, {Langer}, {Castro}, \&
  {Fossati}}]{2015A&A...584A..54P}
{Petermann}, I., {Langer}, N., {Castro}, N., \& {Fossati}, L. 2015, \aap, 584,
  A54

\bibitem[{{Petruk} {et~al.}(2011){Petruk}, {Orlando}, {Beshley}, \&
  {Bocchino}}]{2011MNRAS.413.1657P}
{Petruk}, O., {Orlando}, S., {Beshley}, V., \& {Bocchino}, F. 2011, \mnras,
  413, 1657

\bibitem[{{Petruk} {et~al.}(2017){Petruk}, {Orlando}, {Miceli}, \&
  {Bocchino}}]{2017A&A...605A.110P}
{Petruk}, O., {Orlando}, S., {Miceli}, M., \& {Bocchino}, F. 2017, \aap, 605,
  A110

\bibitem[{{Potter} {et~al.}(2014){Potter}, {Staveley-Smith}, {Reville}, {Ng},
  {Bicknell}, {Sutherland}, \& {Wagner}}]{2014ApJ...794..174P}
{Potter}, T.~M., {Staveley-Smith}, L., {Reville}, B., {et~al.} 2014, \apj, 794,
  174

\bibitem[{{Pumo} \& {Zampieri}(2011)}]{2011ApJ...741...41P}
{Pumo}, M.~L. \& {Zampieri}, L. 2011, Astrophys. J., 741, 41

\bibitem[{{Rembiasz} {et~al.}(2016{\natexlab{a}}){Rembiasz}, {Guilet},
  {Obergaulinger}, {Cerd{\'a}-Dur{\'a}n}, {Aloy}, \&
  {M{\"u}ller}}]{2016MNRAS.460.3316R}
{Rembiasz}, T., {Guilet}, J., {Obergaulinger}, M., {et~al.} 2016{\natexlab{a}},
  \mnras, 460, 3316

\bibitem[{{Rembiasz} {et~al.}(2016{\natexlab{b}}){Rembiasz}, {Obergaulinger},
  {Cerd{\'a}-Dur{\'a}n}, {M{\"u}ller}, \& {Aloy}}]{2016MNRAS.456.3782R}
{Rembiasz}, T., {Obergaulinger}, M., {Cerd{\'a}-Dur{\'a}n}, P., {M{\"u}ller},
  E., \& {Aloy}, M.~A. 2016{\natexlab{b}}, \mnras, 456, 3782

\bibitem[{{Reynolds}(1998)}]{1998ApJ...493..375R}
{Reynolds}, S.~P. 1998, \apj, 493, 375

\bibitem[{{Sana} {et~al.}(2012){Sana}, {de Mink}, {de Koter}, {Langer},
  {Evans}, {Gieles}, {Gosset}, {Izzard}, {Le Bouquin}, \&
  {Schneider}}]{2012Sci...337..444S}
{Sana}, H., {de Mink}, S.~E., {de Koter}, A., {et~al.} 2012, Science, 337, 444

\bibitem[{{Sironi} {et~al.}(2013){Sironi}, {Spitkovsky}, \&
  {Arons}}]{2013ApJ...771...54S}
{Sironi}, L., {Spitkovsky}, A., \& {Arons}, J. 2013, \apj, 771, 54

\bibitem[{{Smith} {et~al.}(2001){Smith}, {Brickhouse}, {Liedahl}, \&
  {Raymond}}]{2001ApJ...556L..91S}
{Smith}, R.~K., {Brickhouse}, N.~S., {Liedahl}, D.~A., \& {Raymond}, J.~C.
  2001, \apjl, 556, L91

\bibitem[{{Storey} \& {Manchester}(1987)}]{1987Natur.329..421S}
{Storey}, M.~C. \& {Manchester}, R.~N. 1987, \nat, 329, 421

\bibitem[{{Sugerman} {et~al.}(2005){Sugerman}, {Crotts}, {Kunkel}, {Heathcote},
  \& {Lawrence}}]{2005ApJS..159...60S}
{Sugerman}, B.~E.~K., {Crotts}, A.~P.~S., {Kunkel}, W.~E., {Heathcote}, S.~R.,
  \& {Lawrence}, S.~S. 2005, Astrophys. J. Suppl. Ser., 159, 60

\bibitem[{{Telezhinsky} {et~al.}(2012){Telezhinsky}, {Dwarkadas}, \&
  {Pohl}}]{2012APh....35..300T}
{Telezhinsky}, I., {Dwarkadas}, V.~V., \& {Pohl}, M. 2012, Astroparticle
  Physics, 35, 300

\bibitem[{{Townsend} \& {Owocki}(2005)}]{2005MNRAS.357..251T}
{Townsend}, R.~H.~D. \& {Owocki}, S.~P. 2005, \mnras, 357, 251

\bibitem[{{Townsend} {et~al.}(2005){Townsend}, {Owocki}, \&
  {Groote}}]{2005ApJ...630L..81T}
{Townsend}, R.~H.~D., {Owocki}, S.~P., \& {Groote}, D. 2005, \apjl, 630, L81

\bibitem[{{ud-Doula} {et~al.}(2008){ud-Doula}, {Owocki}, \&
  {Townsend}}]{2008MNRAS.385...97U}
{ud-Doula}, A., {Owocki}, S.~P., \& {Townsend}, R.~H.~D. 2008, \mnras, 385, 97

\bibitem[{{ud-Doula} {et~al.}(2013){ud-Doula}, {Sundqvist}, {Owocki}, {Petit},
  \& {Townsend}}]{2013MNRAS.428.2723U}
{ud-Doula}, A., {Sundqvist}, J.~O., {Owocki}, S.~P., {Petit}, V., \&
  {Townsend}, R.~H.~D. 2013, \mnras, 428, 2723

\bibitem[{{Wang} {et~al.}(2003){Wang}, {Baade}, {H{\"o}flich}, {Khokhlov},
  {Wheeler}, {Kasen}, {Nugent}, {Perlmutter}, {Fransson}, \&
  {Lundqvist}}]{2003ApJ...591.1110W}
{Wang}, L., {Baade}, D., {H{\"o}flich}, P., {et~al.} 2003, \apj, 591, 1110

\bibitem[{{Wang} {et~al.}(2004){Wang}, {Baade}, {H{\"o}flich}, {Wheeler},
  {Kawabata}, \& {Nomoto}}]{2004ApJ...604L..53W}
{Wang}, L., {Baade}, D., {H{\"o}flich}, P., {et~al.} 2004, \apjl, 604, L53

\bibitem[{{Wang} \& {Wheeler}(2008)}]{2008ARA&A..46..433W}
{Wang}, L. \& {Wheeler}, J.~C. 2008, \araa, 46, 433

\bibitem[{{West} {et~al.}(1987){West}, {Lauberts}, {Schuster}, \&
  {Jorgensen}}]{1987A&A...177L...1W}
{West}, R.~M., {Lauberts}, A., {Schuster}, H.-E., \& {Jorgensen}, H.~E. 1987,
  \aap, 177, L1

\bibitem[{{Wongwathanarat} {et~al.}(2015){Wongwathanarat}, {M{\"u}ller}, \&
  {Janka}}]{2015A&A...577A..48W}
{Wongwathanarat}, A., {M{\"u}ller}, E., \& {Janka}, H.-T. 2015, \aap, 577, A48

\bibitem[{{Woosley}(1988)}]{1988ApJ...330..218W}
{Woosley}, S.~E. 1988, \apj, 330, 218

\bibitem[{{Zanardo} {et~al.}(2010){Zanardo}, {Staveley-Smith}, {Ball},
  {Gaensler}, {Kesteven}, {Manchester}, {Ng}, {Tzioumis}, \&
  {Potter}}]{2010ApJ...710.1515Z}
{Zanardo}, G., {Staveley-Smith}, L., {Ball}, L., {et~al.} 2010, \apj, 710, 1515

\bibitem[{{Zanardo} {et~al.}(2018){Zanardo}, {Staveley-Smith}, {Gaensler},
  {Indebetouw}, {Ng}, {Matsuura}, \& {Tzioumis}}]{2018ApJ...861L...9Z}
{Zanardo}, G., {Staveley-Smith}, L., {Gaensler}, B.~M., {et~al.} 2018, \apjl,
  861, L9

\bibitem[{{Zhekov} {et~al.}(2009){Zhekov}, {McCray}, {Dewey}, {Canizares},
  {Borkowski}, {Burrows}, \& {Park}}]{2009ApJ...692.1190Z}
{Zhekov}, S.~A., {McCray}, R., {Dewey}, D., {et~al.} 2009, \apj, 692, 1190

\end{thebibliography}

\end{document}